\documentclass[twoside,english,sort&compress]{revtex4-1}
\usepackage[T1]{fontenc}
\usepackage[latin9]{inputenc}
\usepackage{geometry}
\usepackage{color}
\usepackage{units}
\usepackage{amstext}
\usepackage{graphicx}
\usepackage{esint}

\makeatletter

\providecommand{\tabularnewline}{\\}

\newcommand{\eqref}[1]{(\ref{#1})}

\usepackage{babel}

\makeatother

\usepackage{babel}
\begin{document}
\title{Spontaneous emission and energy shifts of a Rydberg rubidium atom
close to an optical nanofiber }
\author{E. Stourm$^{1}$, M. Lepers$^{2}$, J. Robert$^{1}$, S. Nic Chormaic$^{3}$,
K. Mølmer$^{4}$, E. Brion$^{5}$ }
\email{etienne.brion@irsamc.ups-tlse.fr}

\address{$^{1}$Université Paris-Saclay, CNRS, Laboratoire de physique des
gaz et des plasmas, 91405, Orsay, France.~\\
$^{2}$Laboratoire Interdisciplinaire Carnot de Bourgogne, CNRS, Université
de Bourgogne Franche-Comté, 21078 Dijon, France.~\\
$^{3}$Light-Matter Interactions for Quantum Technologies Unit, Okinawa
Institute of Science and Technology Graduate University, Onna, Okinawa,
904 - 0495, Japan.~\\
$^{4}$Department of Physics and Astronomy, Aarhus University, Ny
Munkegade 120, DK-8000 Aarhus C, Denmark.~\\
$^{5}$Laboratoire Collisions Agrégats Réactivité, IRSAMC \& UMR5589
du CNRS, Université de Toulouse III Paul Sabatier, F-31062 Toulouse
Cedex 09, France.}
\begin{abstract}
In this paper, we report on numerical\textcolor{red}{{} }calculations
of the spontaneous emission rates and Lamb shifts of a $^{87}\text{Rb}$
atom in a Rydberg-excited state $\left(n\leq30\right)$ located close
to a silica optical nanofiber. We investigate how these quantities
depend on the fiber's radius, the distance of the atom to the fiber,
the direction of the atomic angular momentum polarization as well
as the different atomic quantum numbers. We also study the contribution
of quadrupolar transitions, which may be substantial for highly polarizable
Rydberg states. Our calculations are performed in the macroscopic
quantum electrodynamics formalism, based on the dyadic Green's function
method. This allows us to take dispersive and absorptive characteristics
of silica into account; this is of major importance since Rydberg
atoms emit along many different transitions whose frequencies cover
a wide range of the electromagnetic spectrum. Our work is an important
initial step towards building a Rydberg atom-nanofiber interface for
quantum optics and quantum information purposes. 
\end{abstract}
\maketitle

\section{Introduction}

Within the last two decades, the strong dipole-dipole interaction
experienced by two neighbouring Rydberg-excited atoms \citep{GA94}
has become the main ingredient for many atom-based quantum information
protocol proposals \citep{SWM10}. This interaction can be so large
as to forbid the simultaneous resonant excitation of two atoms if
their separation is less than a specific distance, called the blockade
radius \citep{TFS04}, which typically depends on the intensity of
the laser excitation and the interaction between the Rydberg atoms
\citep{LWR09}. The discovery of this ``Rydberg blockade'' phenomenon
\citep{LFC01,TFS04,SRA04,CRB05,AVG98,VVZ06} paved the way for a new
encoding scheme using atomic ensembles as collective quantum registers
\citep{LFC01,BMS07,BMM07,BPS08} and repeaters \citep{BCA12,ZMH10,HHH10}.

Scalability is one of the crucial requirements for quantum devices
\citep{DiV00} and interfacing atomic ensembles into a quantum network
is a possible way to reach this goal. Photons naturally appear as
ideal information carriers and the photon-based protocols considered
so far include free-space \citep{PM09}, or guided propagation through
optical fibers \citep{BCA12}. The former has the advantage of being
relatively easy to implement, but presents the drawback of strong
losses. The latter requires a cavity quantum electrodynamics setup,
which is experimentally more involved. An alternative option would
be to use optical nanofibers. Such fibers have recently received much
attention \citep{SGH17,NGN16} because the coupling to the evanescent
guided modes of a nanofiber allows for easy-to-implement atom trapping
\citep{BHK04,KBH04,VRS10} and detection \citep{NMM07,DWM09,KSB17}.
This coupling increases in strength as the fiber diameter reduces
and the atoms approach the fiber surface. It has also been shown that
energy could be exchanged between two distant atoms via the guided
modes of the fiber \citep{KDN05}. This suggests that optical nanofibers
could play the role of a communication channel between the nodes of
an atomic quantum network consisting of Rydberg-excited atomic ensembles.

In the perspective of building a quantum network based on Rydberg-blockaded
atomic ensembles linked via an optical nanofiber, we recently studied
the spontaneous emission of a highly-excited (Rydberg) sodium atom
in the neighbourhood of an optical nanofiber made of silica \citep{SZL19}.
To be more specific, we investigated how the atomic emission rates
into the guided and radiative fiber modes are influenced by the radius
of the fiber, the distance of the atom to the fiber and the symmetry
of the Rydberg state. In the spirit of Ref. \citep{KDB05}, we used
the so-called mode function description of the nanofiber which does
not allow one to take absorption and dispersion of the fiber into
account. This point is critical with highly excited atoms since they
can de-excite along many transitions of different frequencies for
which the fiber index is different and potentially complex. This forced
us, in Ref. \citep{SZL19}, to restrict ourselves to Rydberg levels
of moderate principal numbers so that the frequencies of the transitions
involved remain in a nondispersive and nonabsorptive window of the
silica spectrum. By contrast, here, we resort to the framework of
macroscopic quantum electrodynamics based on the dyadic Green's function
\citep{NH12,Buh12}. This formalism enables us to take the exact refractive
index of silica into account and relaxes all constraints on the transitions
we can address. This framework also offers a natural way to compute
not only spontaneous emission rates, but also Lamb shifts and (resonant
and nonresonant) electromagnetic forces the atom is subject to.

In this article, we present the numerical results we obtained with
this approach for a rubidium atom prepared in a Rydberg-excited state
$\left|n\leq30;L=S,P,D;JFM_{F}\right\rangle $ in the vicinity of
a multimode silica optical nanofiber. We chose $^{87}\text{Rb}$ as
it is commonly used in Rydberg atom experiments, like in the recent
experimental work on Rydberg generation next to a nanofiber \citep{RRK19}.
In particular, we show that a non-negligible fraction of spontaneously
emitted light is guided along the fiber and study how it depends on
principal quantum number, $n$, the radius of the nanofiber, $a$,
the distance of the atom to the nanofiber axis, $R$, and the direction
of angular momentum polarization. Interestingly, when the quantum
and fiber axes do not coincide, spontaneous emission becomes directional,
as already noticed for low-excited atoms \citep{KR14,SBC15} due to
the peculiar polarization structure of the field in the neighbourhood
of the fiber. As shown by our calculations, this effect is particularly
strong for photons emitted into the fiber-guided modes and persists
even for high principal quantum numbers, $n$. This is promising in
view of potential applications in chiral quantum information protocols
\citep{LMS17} based on a Rydberg-atom-nanofiber interface. We also
address Lamb shifts and associated dispersion forces that arise. In
particular, we show that, as $n$ increases, the contribution of quadrupolar
transitions becomes more and more important. This contrasts with spontaneous
emission rates for which quadrupolar transitions have negligible influence.

The article is organized as follows. In Sec. \ref{SECGeneral} we
present the system and introduce the important formulae used in our
calculations. In Sec. \ref{SECNumericalResults} we present and interpret
our numerical results for spontaneous emission rates, Lamb shifts
and forces. We conclude in Sec. \ref{SECConclusion} and give perspectives
of our work. More technical details of our work can be found in Appendices.

\section{System and methods\label{SECGeneral}}

\begin{figure}
\begin{centering}
\includegraphics[width=14cm]{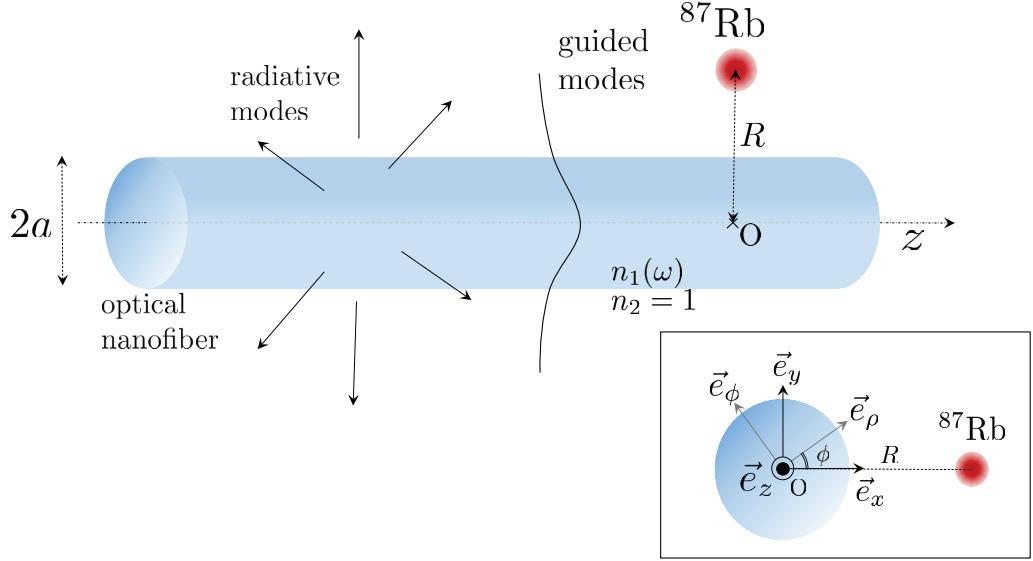} 
\par\end{centering}
\caption{A $^{87}\text{Rb}$ atom located at a distance, $R$, from the axis
of an optical nanofiber of radius, $a$. The refractive index $n_{1}\left(\omega\right)$
for silica is obtained by a numerical fit of the experimental data
taken from \citep{Pal98}. Outside the fiber, the refractive index
is $n_{2}=1$. The axis of the nanofiber is arbitrarily chosen as
the $z$-axis. The cylindrical coordinates $\left(\rho,\phi,z\right)$
and frame $\left(\vec{e}_{\rho},\vec{e}_{\phi},\vec{e}_{z}\right)$
are introduced in the inset.}
\label{System} 
\end{figure}

In this article, we consider a rubidium atom, $^{87}\text{Rb}$, initially
prepared in a highly-excited (Rydberg) level $n\leq30$, located at
a distance $R$ from the axis of a silica nanofiber of radius $a$.
Our goal is to investigate how the fiber modifies the atomic spontaneous
emission rates, the Lamb shifts, and the forces on the atom. To be
more specific, we want to study the influence of: i) the radius of
the fiber, ii) the distance of the atom to the fiber, iii) the different
quantum numbers of the Rydberg state $\left|nLJFM_{F}\right\rangle $,
in particular the principal quantum number $n$, and iv) the direction
of angular momentum polarization on these properties. On Fig. \ref{System},
we define the reference frame $\left(Oxyz\right)$ and the associated
unitary basis $\left(\vec{e}_{x},\vec{e}_{y},\vec{e}_{z}\right)$.
The origin $O$ is chosen as the projection of the atomic center of
mass onto the fiber axis, the $z$-axis coincides with the fiber axis,
and the $x$-axis joins the origin $O$ and the center of mass of
the atom. In this basis, the position vector of the atom is $\vec{R}=R\vec{e}_{x}$.
For future reference we also introduce the cylindrical basis $\left(\vec{e}_{\rho},\vec{e}_{\phi},\vec{e}_{z}\right)$
on Fig. \ref{System}, defined by $\vec{e}_{\rho}=\cos\phi\vec{e}_{x}+\sin\phi\vec{e}_{y}$,
$\vec{e}_{\phi}=-\sin\phi\vec{e}_{x}+\cos\phi\vec{e}_{y}$.

We shall resort to the theoretical framework of macroscopic quantum
electrodynamics \citep{NH12,Buh12}, which allows one to consider
the exact frequency-dependent form of the electric susceptibility
of silica, obtained through a fit of experimental data given in \citep{Pal98}.
This formalism is based on the dyadic Green's function $\overline{\overline{G}}\left(\vec{r},\vec{r}',\omega\right)$,
which is the solution to the Helmholtz equation 
\begin{equation}
\left[\vec{\nabla}\times\vec{\nabla}\times-\varepsilon\left(\vec{r},\omega\right)\frac{\omega^{2}}{c^{2}}\right]\overline{\overline{G}}\left(\vec{r},\vec{r}',\omega\right)=\delta\left(\vec{r}-\vec{r}'\right)\overline{\overline{I}},\label{Helmholtz}
\end{equation}
where $\varepsilon\left(\vec{r},\omega\right)$ is the relative electric
permittivity of the medium at the position $\vec{r}$ and frequency
$\omega$ while $\overline{\overline{I}}$ is the unit dyadic \citep{Tai94}.
The solution of Eq. (\ref{Helmholtz}) in the case of a cylindrical
nanofiber is given in Appendix \ref{AppGDF}. There exist two useful
decompositions of $\overline{\overline{G}}$ : i) $\overline{\overline{G}}=\overline{\overline{G}}_{0}+\overline{\overline{G}}_{\text{sc}}$
where $\overline{\overline{G}}_{0}$ is the vacuum component, and
$\overline{\overline{G}}_{\text{sc}}$ the scattering contribution
due to the presence of the nanofiber and ii) $\overline{\overline{G}}=\overline{\overline{G}}_{\text{g}}+\overline{\overline{G}}_{\text{r}}$
where $\overline{\overline{G}}_{\text{g,r}}$ are the respective contributions
of the guided and radiative modes.

We summarize below the main formulae we used to obtain the results
presented in the next section, the derivation of which can be found
in \citep{Buh12,BKW04}. The spontaneous emission rate, $\Gamma_{n}$,
from an excited state, $\left|n\right\rangle $, is given by the sum,
$\Gamma_{n}=\sum_{k<n}\Gamma_{nk}$, of rates

\begin{eqnarray}
\Gamma_{nk} & = & \frac{2\mu_{0}}{\hbar}\omega_{nk}^{2}\vec{d}_{nk}\cdot\mathrm{Im}\left[\overline{\overline{G}}\left(\vec{R},\vec{R},\omega_{nk}\right)\right]\cdot\vec{d}_{kn}\label{EqGammank}
\end{eqnarray}
relative to the different transitions $\left|n\right\rangle \rightarrow\left|k\right\rangle $
for $k<n$, where $\omega_{nk}$ and $\vec{d}_{nk}\equiv\left\langle n\left|\hat{\vec{d}}\right|k\right\rangle $
denote the bare frequency and the dipole matrix element of the transition
$\left|k\right\rangle \rightarrow\left|n\right\rangle $, respectively.

In the same way, the Lamb shift, $\delta\omega_{n}$, of an excited
state, $\left|n\right\rangle $, is given by the sum, $\delta\omega_{n}=\sum_{k}\delta\omega_{nk}$,
of all energy shifts induced by the different transitions $\left|n\right\rangle \rightarrow\left|k\right\rangle $,
for arbitrary $k\neq n$, with 
\begin{eqnarray}
\delta\omega_{nk} & = & -\frac{\mu_{0}}{\hbar\pi}\mathcal{P}\left(\intop_{0}^{+\infty}\mathrm{d}\omega~\frac{\omega^{2}}{\omega-\omega_{nk}}\vec{d}_{nk}\cdot\mathrm{Im}\left[\overline{\overline{G}}\left(\vec{R},\vec{R},\omega\right)\right]\cdot\vec{d}_{kn}\right)\label{EqLS}
\end{eqnarray}
where $\mathcal{P}$ denotes the Cauchy principal value. Here, we
shall use the non-retarded approximation \citep{EBS11} 
\begin{eqnarray}
\delta\omega_{nk} & \approx & -\frac{1}{2\hbar\epsilon_{0}}\vec{d}_{nk}\cdot\overline{\overline{\Gamma}}_{0}\left(\vec{R}\right)\cdot\vec{d}_{kn}\label{EqLSNR}
\end{eqnarray}
where $\overline{\overline{\Gamma}}_{0}\left(\vec{R}\right)=\lim_{\omega\rightarrow0}\frac{\omega^{2}}{c^{2}}\overline{\overline{G}}\left(\vec{R},\vec{R},\omega\right)$.
This approximation is particularly suited for Rydberg atoms, since
the main contributions to the Lamb shift are due to transitions to
neighbouring states, therefore of long wavelengths.

Finally, the average resonant and nonresonant forces on an atom initially
in the state $\left|n\right\rangle $, evaluated at $t=0$, are given
by (see Appendix \ref{AppF})

\begin{eqnarray}
\vec{F}^{\text{res}}\left(t=0\right) & = & \sum_{k}\left[\mu_{0}\omega_{nk}^{2}\vec{\nabla}_{\vec{r}}\left.\left[\vec{d}_{nk}\cdot\overline{\overline{G}}_{\text{sc}}\left(\vec{r},\vec{R},\omega_{nk}\right)\cdot\vec{d}_{kn}\right]\right|_{\vec{r}=\vec{R}}+\text{c.c.}\right]\label{EqFRes}\\
\vec{F}^{\text{nonres}}\left(t=0\right) & = & -\frac{\mu_{0}}{\pi}\intop_{0}^{+\infty}\mathrm{d}\xi~\xi^{2}\frac{\omega_{kn}}{\omega_{kn}^{2}+\xi^{2}}\nabla_{\vec{r}}\left[\vec{d}_{nk}\cdot\overline{\overline{G}}_{\text{sc}}\left(\vec{r},\vec{R},\mathrm{i}\xi\right)|_{\vec{r}=\vec{R}}\cdot\vec{d}_{kn}\right].\label{EqFNRes}
\end{eqnarray}
where $\vec{\nabla}_{\vec{r}}$ acts on the spatial variable, $\vec{r}$.

\section{Numerical results and discussion \label{SECNumericalResults}}

In this section we present and interpret the numerical results we
obtained for spontaneous emission rates and Lamb shifts of a $^{87}\text{Rb}$
atom in the vicinity of a silica optical nanofiber. In particular,
we investigate the effect of the distance, $R$, from the atom to
the fiber axis, the fiber radius, $a$, and the atomic quantum numbers.
We also study the influence of the direction of angular momentum polarization
on the strength and directionality of spontaneous emission from a
Rydberg level, specifically towards the guided modes. Finally, we
address quadrupolar transitions, which, \emph{a priori}, may have
a substantial influence on Rydberg atom emission properties in view
of their high polarizability.

\subsection{Spontaneous emission rates}

We start the discussion with the results we obtained for spontaneous
emission rates. In Secs. \ref{DepGamR}-\ref{GamQuad}, the quantization
axis is implicitly chosen along the fiber axis $\left(Oz\right)$.
In contrast, in Secs. \ref{DepGamQuantAx}-\ref{GamAnis}, we investigate
the changes induced by other quantization axis choices. In some places,
for pedagogical reasons, we shall resort to the so-called mode function
approach (widely used in the works by F. Le Kien, see, e.g. \citep{KBH04})
as it offers a simple and illustrative way to physically interpret
our results. However, we wish to emphasise that our calculations were
performed using the (more general) Green's function formalism, which
allows one to account for dispersive and absorptive characteristics
of the fiber.

\subsubsection{Dependence on the distance, $R$, from the atom to the fiber axis
\label{DepGamR}}

\begin{figure}
\begin{centering}
\includegraphics[width=16cm]{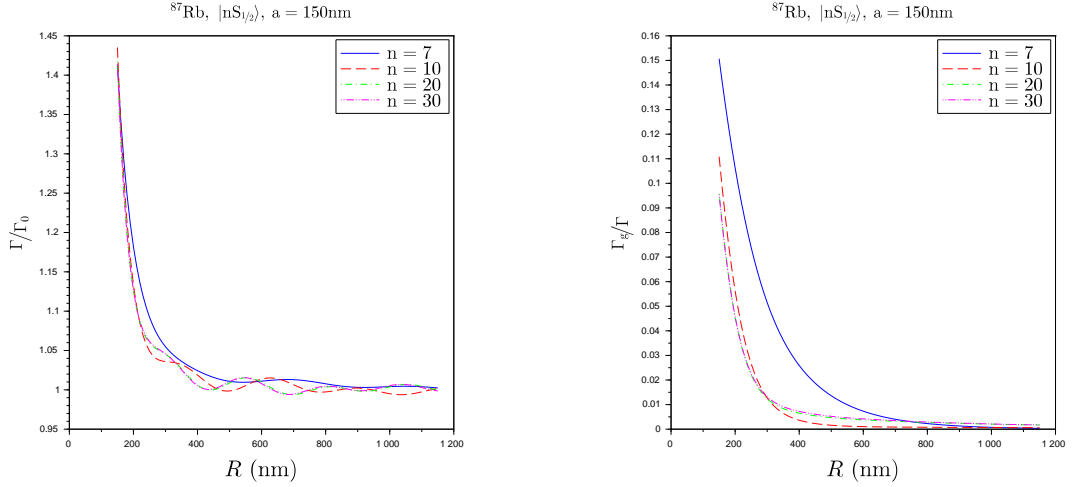} 
\par\end{centering}
\caption{\textbf{Spontaneous emission rates of an $^{87}\text{Rb}$ atom in
the state }$\left|nS_{\nicefrac{1}{2}}\right\rangle $\textbf{ (with
$n=7,10,20,30$) -- dependence on the distance, $R$, from the atom
to the nanofiber}. We represent the ratios $\nicefrac{\Gamma}{\Gamma_{0}}$
(left), $\nicefrac{\Gamma_{g}}{\Gamma}$ (right) as functions of $R$.
$\Gamma_{g}$ and $\Gamma_{r}$ denote the spontaneous emission rates
towards the guided and radiative modes, respectively, $\Gamma\equiv\Gamma_{g}+\Gamma_{r}$
is the total spontaneous emission rate and $\Gamma_{0}$ the spontaneous
emission rate in vacuum. The radius of the nanofiber is fixed at $a=150$
nm. }
\label{FigDepRS} 
\end{figure}

\begin{figure}
\begin{centering}
\includegraphics[width=16cm]{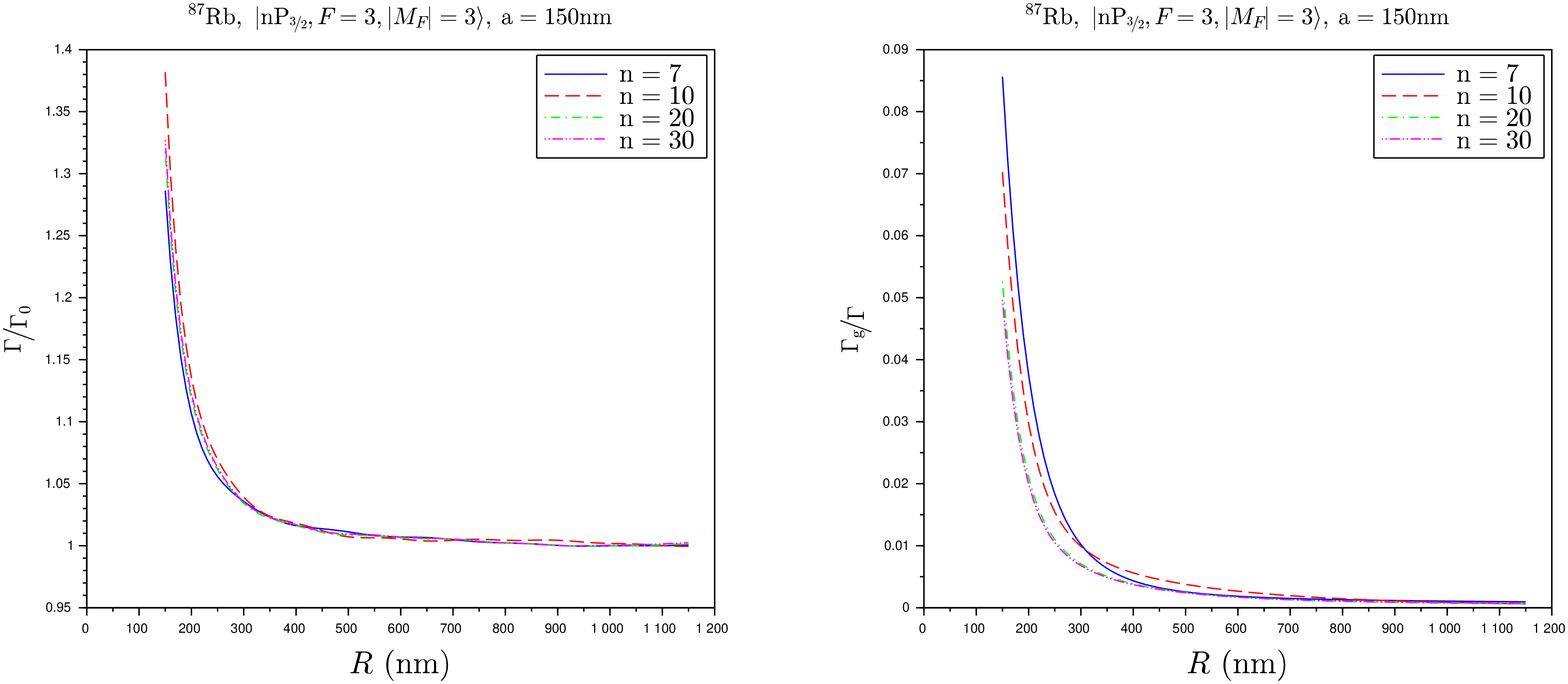} 
\par\end{centering}
\caption{\textbf{Spontaneous emission rates of an $^{87}\text{Rb}$ atom in
the state }$\left|nP_{\nicefrac{3}{2}},F=3,M_{F}=3\right\rangle $\textbf{
(with $n=7,10,20,30$) -- dependence on the distance, $R$, from
the atom to the nanofiber}. We represent the ratios $\nicefrac{\Gamma}{\Gamma_{0}}$
(left), $\nicefrac{\Gamma_{g}}{\Gamma}$ (right) as functions of $R$.
$\Gamma_{g}$, $\Gamma_{r}$ denote the spontaneous emission rates
into the guided and radiative modes, respectively, $\Gamma\equiv\Gamma_{g}+\Gamma_{r}$
is the total spontaneous emission rate and $\Gamma_{0}$ the spontaneous
emission rate in vacuum. The radius of the nanofiber is fixed at $a=150$~nm. }
\label{FigDepRP} 
\end{figure}

\begin{figure}
\begin{centering}
\includegraphics[width=16cm]{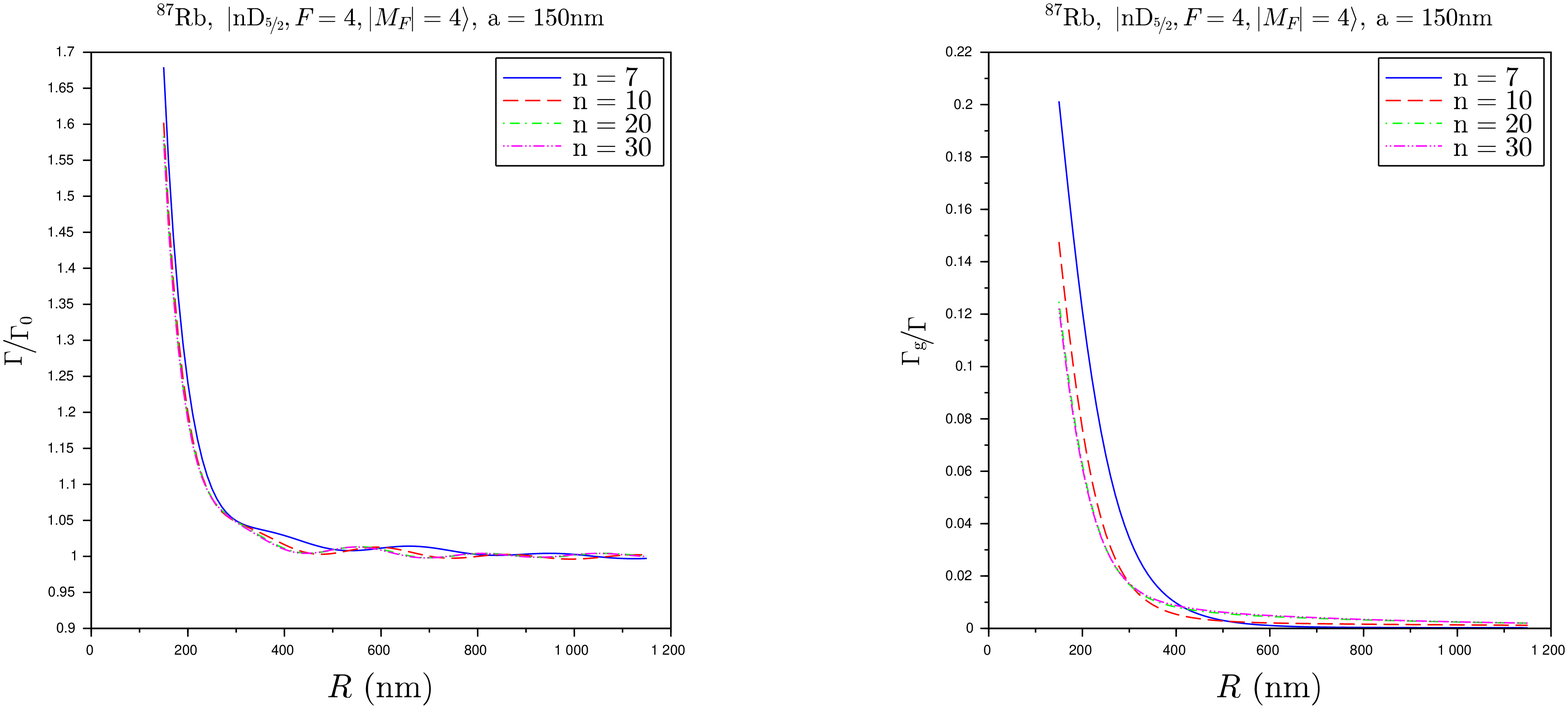} 
\par\end{centering}
\caption{\textbf{Spontaneous emission rates of an $^{87}\text{Rb}$ atom in
the state }$\left|nD_{\nicefrac{5}{2}},F=4,M_{F}=4\right\rangle $\textbf{
(with $n=7,10,20,30$) -- dependence on the distance, $R$, from
the atom to the nanofiber}. We represent the ratios $\nicefrac{\Gamma}{\Gamma_{0}}$
(left), $\nicefrac{\Gamma_{g}}{\Gamma}$ (right) as functions of $R$.
$\Gamma_{g}$, $\Gamma_{r}$ denote the spontaneous emission rates
into the guided and radiative modes, respectively, $\Gamma\equiv\Gamma_{g}+\Gamma_{r}$
is the total spontaneous emission rate and $\Gamma_{0}$ is the spontaneous
emission rate in vacuum. The radius of the nanofiber is fixed at $a=150$~nm. }
\label{FigDepRD} 
\end{figure}

Figures \ref{FigDepRS}, \ref{FigDepRP} and \ref{FigDepRD} show
the variations with the distance, $R$, from the atom to the nanofiber
axis of: i) the ratio $\nicefrac{\Gamma}{\Gamma_{0}}$ of the total
spontaneous emission rate of the atom to the spontaneous emission
rate in vacuum, ii) the ratio $\nicefrac{\Gamma_{g}}{\Gamma}$ of
the spontaneous emission rate of the atom only into the guided modes
to the total spontaneous emission rate for the states $\left|nS_{\nicefrac{1}{2}}\right\rangle $,
$\left|nP_{\nicefrac{3}{2}},F=3,M_{F}=3\right\rangle $, 

$\left|nD_{\nicefrac{5}{2}},F'=4,M_{F'}=4\right\rangle $, respectively,
with $n=7,10,20,30$, and for a nanofiber radius $a=150\text{ nm}$.

\begin{table}
\begin{centering}
\begin{tabular}{|c|c|c|c|c|}
\hline 
$n$  & 7  & 10  & 20  & 30\tabularnewline
\hline 
\hline 
$\left|nS_{\nicefrac{1}{2}}\right\rangle $  & $1.132\times\mathrm{10}^{7}$  & $2.375\times\mathrm{10}^{6}$  & $1.662\times\mathrm{10}^{5}$  & $4.120\times\mathrm{10}^{4}$\tabularnewline
\hline 
$\left|nP_{\nicefrac{1}{2}}\right\rangle $  & $1.624\times\mathrm{10}^{6}$  & $7.424\times\mathrm{10}^{5}$  & $6.252\times\mathrm{10}^{4}$  & $1.624\times\mathrm{10}^{4}$\tabularnewline
\hline 
$\left|nD_{\nicefrac{5}{2}},F=4,M_{F}=4\right\rangle $  & $2.642\times\mathrm{10}^{6}$  & $1.092\times\mathrm{10}^{6}$  & $1.328\times\mathrm{10}^{5}$  & $3.780\times\mathrm{10}^{4}$\tabularnewline
\hline 
\end{tabular}
\par\end{centering}
\caption{Theoretical values of the spontaneous emission rate, $\Gamma_{0}$,
in vacuum of an $^{87}\text{Rb}$ atom in the states $|\mathrm{nS}_{1/2}\rangle$,
$|\mathrm{nP}_{1/2}\rangle$ and $|\mathrm{nD}_{5/2},F=4,M_{F}=4\rangle$
for $n=7,10,20,30$ (in $\text{s}^{-1}$).}

\label{TabGamma0} 
\end{table}

In all cases, close to the nanofiber, the total spontaneous emission
is amplified when compared with its value in vacuum. This amplification
vanishes as $R$ increases. The small Drexhage-like oscillations observed
\citep{D70} are due to the oscillatory behavior of the radiative
modes themselves.

Close to the fiber, a non-negligible fraction of the spontaneous emission
is captured by the guided modes. The strongest effect is obtained
for $S$ and $D$ states, as already noted and interpreted in \citep{SZL19}.
As $R$ increases, the guided modes are (quasi-)exponentially damped,
hence the damping of $\Gamma_{g}$ itself.

The dependence with $n$ is less easy to interpret. Let us first note
that $\Gamma$, $\Gamma_{g}$ and $\Gamma_{0}$ substantially decrease
when the principal quantum number increases (see Table \ref{TabGamma0}
for theoretical values of $\Gamma_{0}$). The ratios $\nicefrac{\Gamma}{\Gamma_{0}}$
and $\nicefrac{\Gamma_{g}}{\Gamma}$, however, keep the same order
of magnitude and, therefore, the plots in Figs. \ref{FigDepRS}, \ref{FigDepRP}
and \ref{FigDepRD} for $n=7,10,20,30$ remain close to each other.
In particular, for high values of $n$, the plots seem to tend to
an asymptotic curve. This observation can be qualitatively understood
as follows. We first note that, for high $n$, only a few transitions
substantially contribute to the spontaneous emission rate. In the
crude but practical two-level approximation, we assume the spontaneous
emission rate is dominated by one transition $\left|n\right\rangle \rightarrow\left|k\right\rangle $
whose total spontaneous emission rate, spontaneous emission rate towards
guided modes and spontaneous emission rate in vacuum are, respectively,
given by 
\begin{eqnarray*}
\Gamma_{nk} & = & \frac{2\mu_{0}}{\hbar}\omega_{nk}^{2}\vec{d}_{nk}\cdot\mathrm{Im}\left[\overline{\overline{G}}\left(\vec{R},\vec{R},\omega_{nk}\right)\right]\cdot\vec{d}_{kn}\\
\Gamma_{\text{g},nk} & = & \frac{2\mu_{0}}{\hbar}\omega_{nk}^{2}\vec{d}_{nk}\cdot\mathrm{Im}\left[\overline{\overline{G}}_{\text{g}}\left(\vec{R},\vec{R},\omega_{nk}\right)\right]\cdot\vec{d}_{kn}\\
\Gamma_{0,nk} & = & \frac{\omega_{nk}^{3}\vec{d}_{nk}^{2}}{3\pi\hbar\varepsilon_{0}c^{3}}.
\end{eqnarray*}
For increasing $n$, $\omega_{nk}$ saturates, i.e., Rydberg levels
are closer and closer in energy as the principal quantum number grows,
and the terms $\omega_{nk}^{2}\mathrm{Im}\left[\overline{\overline{G}}\left(\vec{R},\vec{R},\omega_{nk}\right)\right]$
and $\omega_{nk}^{2}\mathrm{Im}\left[\overline{\overline{G}}_{g}\left(\vec{R},\vec{R},\omega_{nk}\right)\right]$,
therefore, also saturate. Finally, since $\Gamma,\Gamma_{\text{g}},\Gamma_{0}\propto\left|\vec{d}_{nk}\right|^{2}$,
the ratios $\nicefrac{\Gamma}{\Gamma_{0}}$ and $\nicefrac{\Gamma_{g}}{\Gamma}$
do not (substantially) depend on the dipole and saturate as $n$ increases.

\subsubsection{Dependence on the fiber radius, $a$\label{DepGama}}

\begin{figure}
\begin{centering}
\includegraphics[width=16cm]{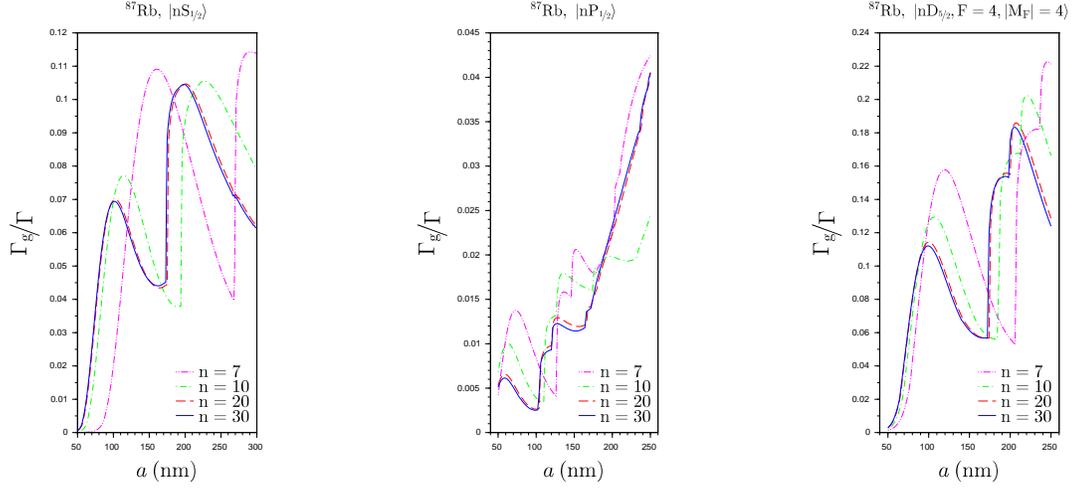} 
\par\end{centering}
\caption{\textbf{Spontaneous emission of an $^{87}\text{Rb}$ atom near an
optical nanofiber -- dependence on the fiber radius, $a$. }We represent
the ratio $\nicefrac{\Gamma_{g}}{\Gamma}$, for an $^{87}\text{Rb}$
atom in the states $\left|nS_{\nicefrac{1}{2}}\right\rangle $ (left),
$\left|nP_{\nicefrac{1}{2}}\right\rangle $ (middle) and $\left|nD_{\nicefrac{5}{2}},F=4,\left|M_{F}\right|=4\right\rangle $
(right), with $n=\left(7,10,20,30\right)$, as a function of $a$.
The atom is located 50~nm from the fiber (i.e., $R=a+$50~nm).}

\label{FigDepaSPD} 
\end{figure}

Figure \ref{FigDepaSPD} shows the dependence on the fiber radius,
$a$, of the ratio $\nicefrac{\Gamma_{g}}{\Gamma}$ for an $^{87}\text{Rb}$
atom in the states $\left|nS_{\nicefrac{1}{2}}\right\rangle $ (left),
$\left|nP_{\nicefrac{1}{2}}\right\rangle $ (middle) and $\left|nD_{\nicefrac{5}{2}},F=4,\left|M_{F}\right|=4\right\rangle $
(right), with $n=\left(7,10,20,30\right)$. The atom is located at
a distance $d=50\text{ nm}$ from the fiber surface, i.e., $R=a+50$~nm
from the fiber axis. Note that the contributions of all guided modes
are summed.

The ratio $\nicefrac{\Gamma_{g}}{\Gamma}$ exhibits the same qualitative
behavior with respect to $a$ for $S$ and $D$ states, and $\left(\nicefrac{\Gamma_{g}}{\Gamma}\right)_{S,D}\approx10\left(\nicefrac{\Gamma_{g}}{\Gamma}\right)_{P}$.
Note that, for the states $\left|nS_{\nicefrac{1}{2}}\right\rangle $
and $\left|nP_{\nicefrac{1}{2}}\right\rangle $, the hyperfine states
(recall $I=\frac{3}{2}$ for $^{87}\text{Rb}$) have the same $\Gamma_{g}$.
This is not the case for $\left|nD_{\nicefrac{5}{2}}\right\rangle $
and in Fig. \ref{FigDepaSPD}, we chose to represent the specific
``edge'' hyperfine state $\left|nD_{\nicefrac{5}{2}},F=4,M_{F}=4\right\rangle $.

The abrupt slope changes observed in all plots originate from the
appearance of additional guided modes as $a$ increases. To be more
explicit, the successive maxima of $\nicefrac{\Gamma_{g}}{\Gamma}$
can be interpreted as follows: i) As a function of the fiber radius,
the amplitude of a specific guided mode at the location of the atom,
i.e. at a distance $d$ from the fiber surface, exhibits a maximum
for a specific value, denoted by $a_{max}\left(\omega,d\right)$,
which depends both on the frequency of the mode and the distance,
$d$. (Note that $a_{max}$ actually also depends on other characteristics
of the mode such as polarization, and wavevector). ii) For a given
atomic transition, of frequency, $\omega_{0}$, the coupling to a
given mode reaches its maximum when $a=a_{max}\left(\omega_{0},d\right)$,
hence a peak in $\nicefrac{\Gamma_{g}}{\Gamma}$.

\begin{figure}
\begin{centering}
\includegraphics[width=16cm]{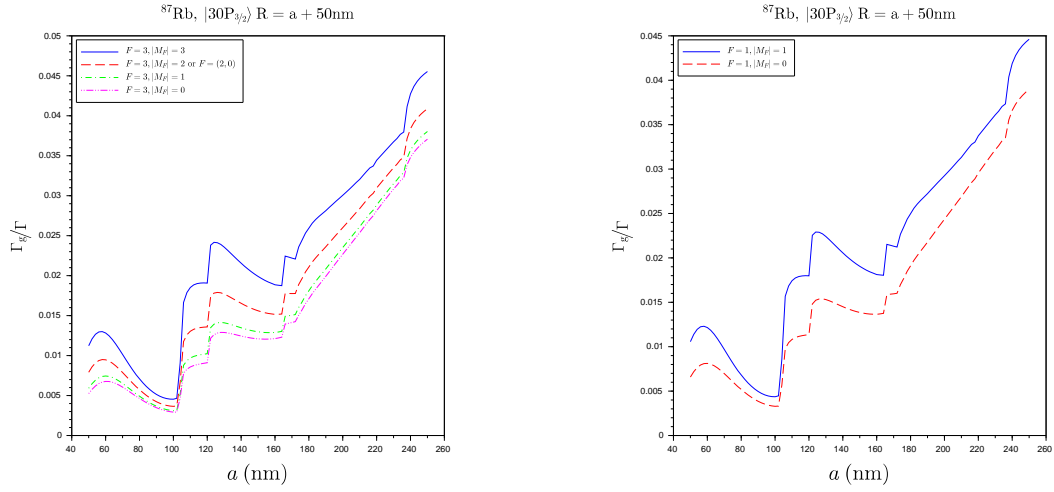} 
\par\end{centering}
\caption{\textbf{Spontaneous emission of an $^{87}\text{Rb}$ atom near an
optical nanofiber -- dependence on the fiber radius, $a$. }We represent
the ratio $\nicefrac{\Gamma_{g}}{\Gamma}$ for an $^{87}\text{Rb}$
atom in the states $\left|30P_{\nicefrac{3}{2}},F=0,2,3,\left|M_{F}\right|=0\cdots F\right\rangle $
(left) and $\left|30P_{\nicefrac{3}{2}},F=1,\left|M_{F}\right|=0\cdots F\right\rangle $
(right) as a function of $a$. The atom is located $50$~nm from
the fiber (i.e., $R=a+50$~nm). }

\label{FigDepaP} 
\end{figure}

Figure \ref{FigDepaP} shows the dependence on the fiber radius, $a$,
of the ratio $\nicefrac{\Gamma_{g}}{\Gamma}$ for an $^{87}\text{Rb}$
atom in the states $\left|30P_{\nicefrac{3}{2}},F=0\cdots3,\left|M_{F}\right|=0\cdots F\right\rangle $
located at a distance $d=50$~nm from the fiber surface, i.e., $R=a+50$~nm
from the fiber axis. As can be observed in the figure, though the
different hyperfine magnetic sublevels for a given $F$ show the same
qualitative behavior, the spontaneous emission towards the guided
modes is stronger for states of higher $\left|M_{F}\right|$. This
can be qualitatively understood as follows: i) Guided modes have a
large (though not exclusive) transverse component, i.e., orthogonal
to the fiber axis $\left(Oz\right)$ (see Fig. \ref{System}); ii)
High coupling to the guided modes is, therefore, obtained for transitions
corresponding to dipoles in the transverse plane $\left(Oxy\right)$;
iii) The quantization axis being along the fiber axis, dipoles in
the plane $\left(Oxy\right)$ correspond to $\sigma$-transitions:
therefore, the stronger the weight of $\sigma$-transitions in the
de-excitation of an excited state, the higher the spontaneous emission
rate towards guided modes; iv) The higher the value of $\left|M_{F}\right|$,
the stronger the weight of $\sigma$-transitions in the de-excitation
of the state (this can be directly checked on $3j$-coefficients),
therefore, the higher $\left|M_{F}\right|$, the higher the spontaneous
emission rate towards guided modes.

\begin{figure}
\begin{centering}
\includegraphics[width=16cm]{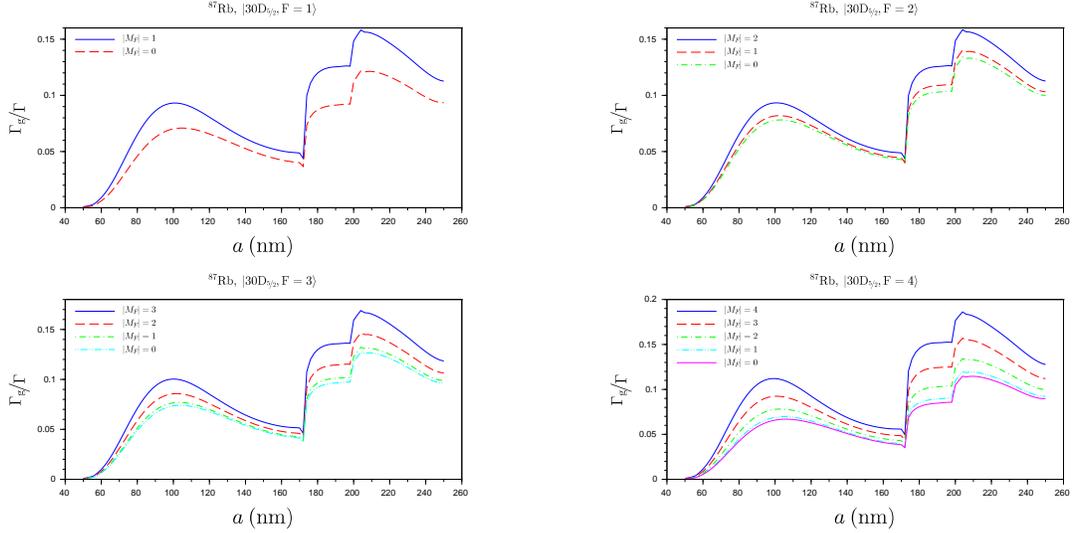} 
\par\end{centering}
\caption{\textbf{Spontaneous emission of an $^{87}\text{Rb}$ atom near an
optical nanofiber -- dependence on the fiber radius, $a$. }We represent
the ratio $\nicefrac{\Gamma_{g}}{\Gamma}$ for an $^{87}\text{Rb}$
atom in the states $\left|30D_{\nicefrac{5}{2}};F=1,\cdots,4;\left|M_{F}\right|=0\cdots F\right\rangle $
as a function of $a$. The atom is located at $50$~nm from the fiber
(i.e., $R=a+50$~nm).}

\label{FigDepaD} 
\end{figure}

The same behavior can be observed and interpreted in Fig. \ref{FigDepaD}
for the states 

$\left|30D_{\nicefrac{5}{2}},F=1\cdots4,\left|M_{F}\right|=0\cdots F\right\rangle $.

\subsubsection{Role of quadrupolar transitions\label{GamQuad}}

Because of their polarizability, Rydberg atoms are very sensitive
to electric fields and electric field inhomogeneities. It is, therefore,
reasonable to expect quadrupolar transitions to play a role in the
de-excitation of a Rydberg atom in the vicinity of an optical nanofiber
where spatial variations of the field are very rapid. Following \citep{KRN18,KD05,CS19},
we calculate the correction due to electric quadrupolar transitions
on the spontaneous emission rates of an $^{87}\text{Rb}$ atom in
the state $\left|nS_{\nicefrac{1}{2}}\right\rangle $ located close
to a silica optical nanofiber (see Appendix \ref{AppEDT} for more
details).

Figure \ref{FigQtot} (left) shows the dependence on $n$ of the electric
quadrupolar transition correction, $\Gamma_{r}^{Q}$, to the spontaneous
emission rate into the radiative modes, for two values of the nanofiber
radius, $a=100$ and 200~nm. To obtain the strongest effect, we fixed
$R=a$, corresponding to the unrealistic situation in which the atom
is located at the fiber surface. As expected, for smaller values of
$a$, the field inhomogeneities are more pronounced and the effect
of electric quadrupolar transitions is higher. Moreover, the contribution
$\Gamma_{r}^{Q}$ decreases with increasing $n$, in the same way
as the coupling to ground states that is responsible for the spontaneous
emission.

The same observations can be made from Fig. \ref{FigQtot} (middle,
right), which show the dependence on $n$ of the electric quadrupolar
transition corrections $\Gamma_{g}^{Q}$ and $\Gamma_{0}^{Q}$ to
the spontaneous emission rate into the first guided modes and vacuum,
respectively. To obtain the strongest effect, we again fixed $R=a$.
We, moreover, note that $\Gamma_{r}^{Q}\gg\Gamma_{g}^{Q}\approx\Gamma_{0}^{Q}$.

Generally speaking, a comparison to the values calculated in the previous
section shows that the quadrupolar contribution is negligible. In
contrast, quadrupolar transitions play an important role in the Lamb
shift, as we shall see below.

\begin{figure}
\begin{centering}
\includegraphics[width=16cm]{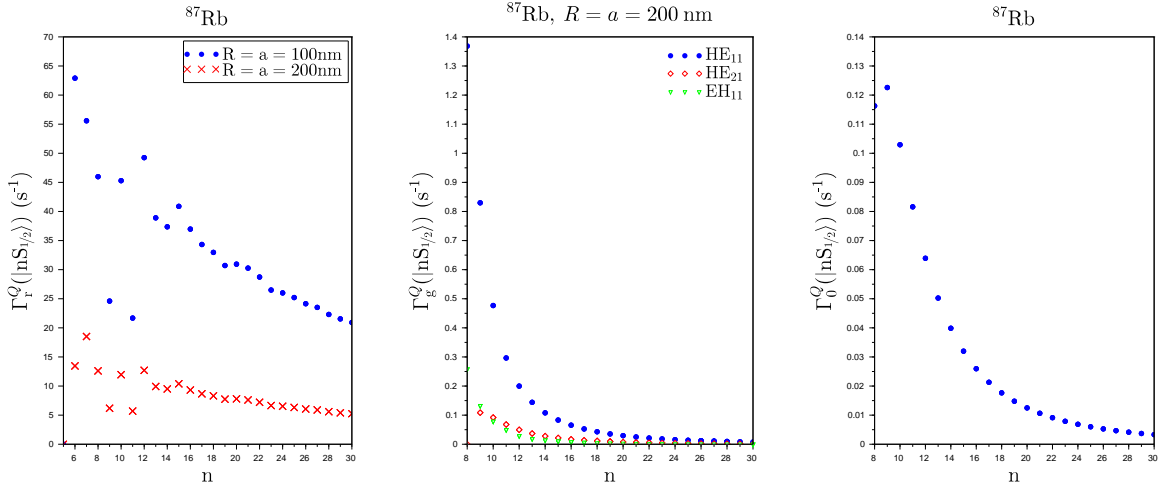} 
\par\end{centering}
\caption{\textbf{Spontaneous emission of an $^{87}\text{Rb}$ atom near an
optical nanofiber -- Contribution of the electric quadrupolar transitions.
} We represent the contribution of the electric quadrupolar transitions
to the spontaneous emission rates into the radiative modes, $\Gamma_{r}^{Q}$
(left), the first guided modes, $\Gamma_{g}^{Q}$ (middle), and in
vacuum, $\Gamma_{0}^{Q}$ (right), for an $^{87}\text{Rb}$ atom in
the state $\left|nS_{\nicefrac{1}{2}}\right\rangle $ as a function
of the principal quantum number, $n$. To get the highest possible
value, we assume the atom is located on the nanofiber surface, i.e.,
$R=a$. In the case of the radiative modes (left), we considered two
values for the fiber radius $a=100$ and $200$~nm, while $a=200$~nm
for the other two plots.}

\label{FigQtot} 
\end{figure}

\subsubsection{Influence of the quantization axis\label{DepGamQuantAx}}

\begin{figure}
\begin{centering}
\includegraphics[width=14cm]{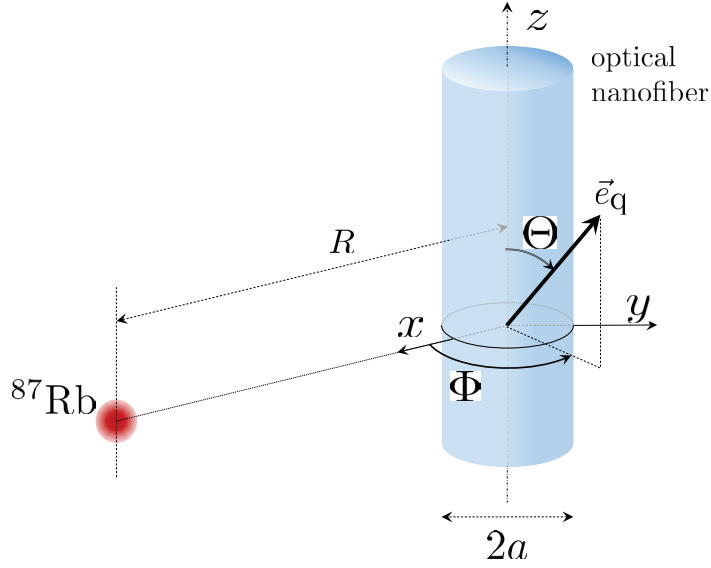} 
\par\end{centering}
\caption{Definition of the angles $\left(\Theta,\Phi\right)$ characterizing
the quantization axis directed along the unitary vector $\vec{e}_{\text{q}}\equiv\sin\Theta\cos\Phi\vec{e}_{x}+\sin\Theta\sin\Phi\vec{e}_{y}+\cos\Theta\vec{e}_{z}$.}
\label{AnglesQuantizationAxis} 
\end{figure}

Until now, the quantization axis was implicitly fixed along the fiber
axis $\left(Oz\right)$. Here, in the spirit of the experimental work
in Ref \citep{SGX19}, we study how the spontaneous emission rate
of an atom close to an optical nanofiber depends on the direction
of the quantization axis chosen to define its state, and therefore
the direction of its angular momentum polarization. The angles $\left(\Theta,\Phi\right)$
characterizing the quantization axis are specified in Fig. \ref{AnglesQuantizationAxis}.

\begin{figure}
\begin{centering}
\includegraphics[width=16cm]{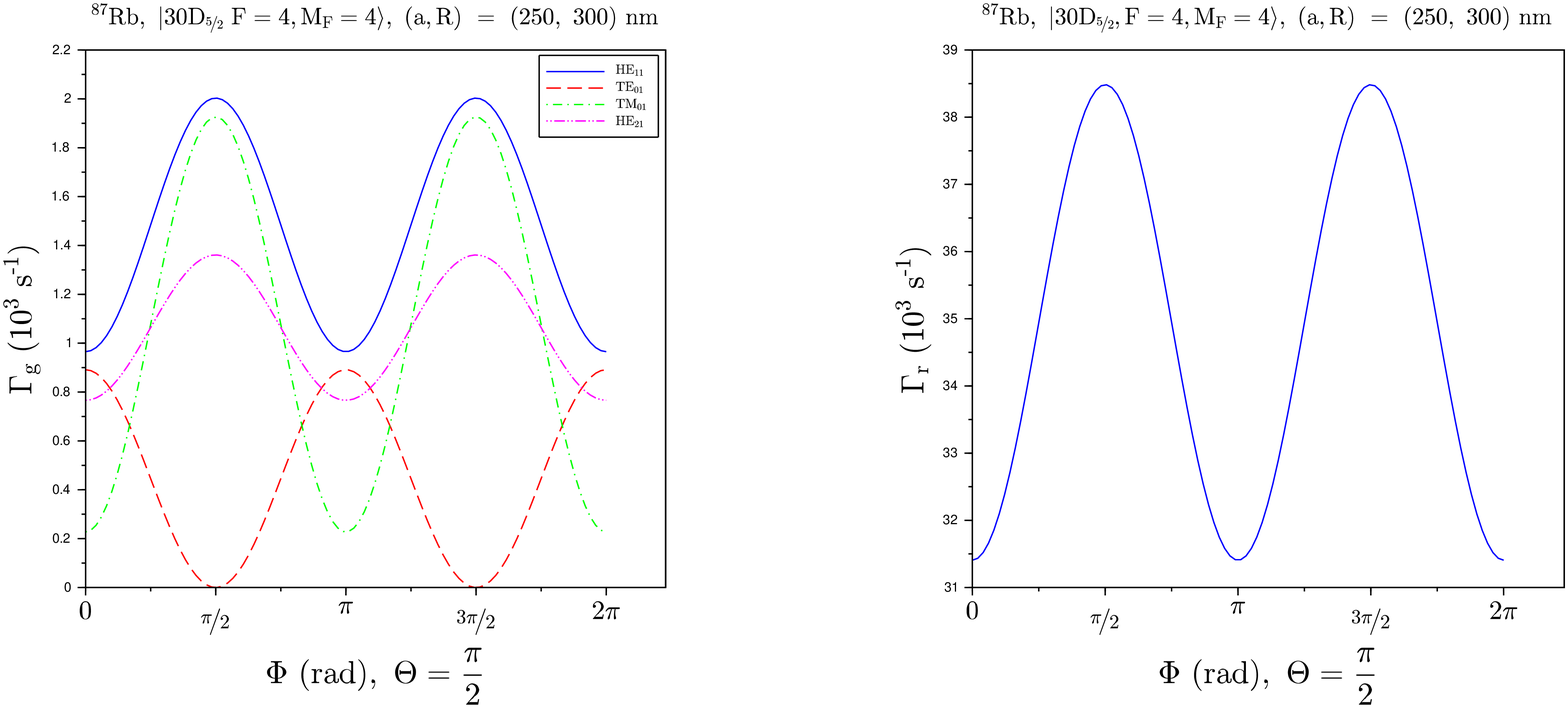} 
\par\end{centering}
\caption{\textbf{Spontaneous emission of an $^{87}\text{Rb}$ atom near an
optical nanofiber with quantization axis in the $\left(Oxy\right)$
plane. }We plot the spontaneous emission rates, $\Gamma_{g}$ (left)
and $\Gamma_{r}$ (right), into the first guided and radiative modes,
respectively, for an $^{87}\text{Rb}$ atom in the state $\left|30D_{\nicefrac{5}{2}},F=4,M_{F}=4\right\rangle $
as functions of the angle $\Phi$ (c.f. Fig. \ref{AnglesQuantizationAxis}),
with $\Theta=\nicefrac{\pi}{2}$. The contributions to $\Gamma_{g}$
of the first four guided modes, $\text{HE}_{11},\text{TE}_{01}$,
$\text{\ensuremath{\text{TM}_{01}}}$ and $\text{\ensuremath{\text{HE}_{21}}}$,
are displayed separately. The radius of the fiber is $a=250$~nm
and the atom is located $50$~nm from the fiber (i.e., $R=a+50$~nm).}

\label{QuantAxOXY} 
\end{figure}

\begin{figure}
\begin{centering}
\includegraphics[width=16cm]{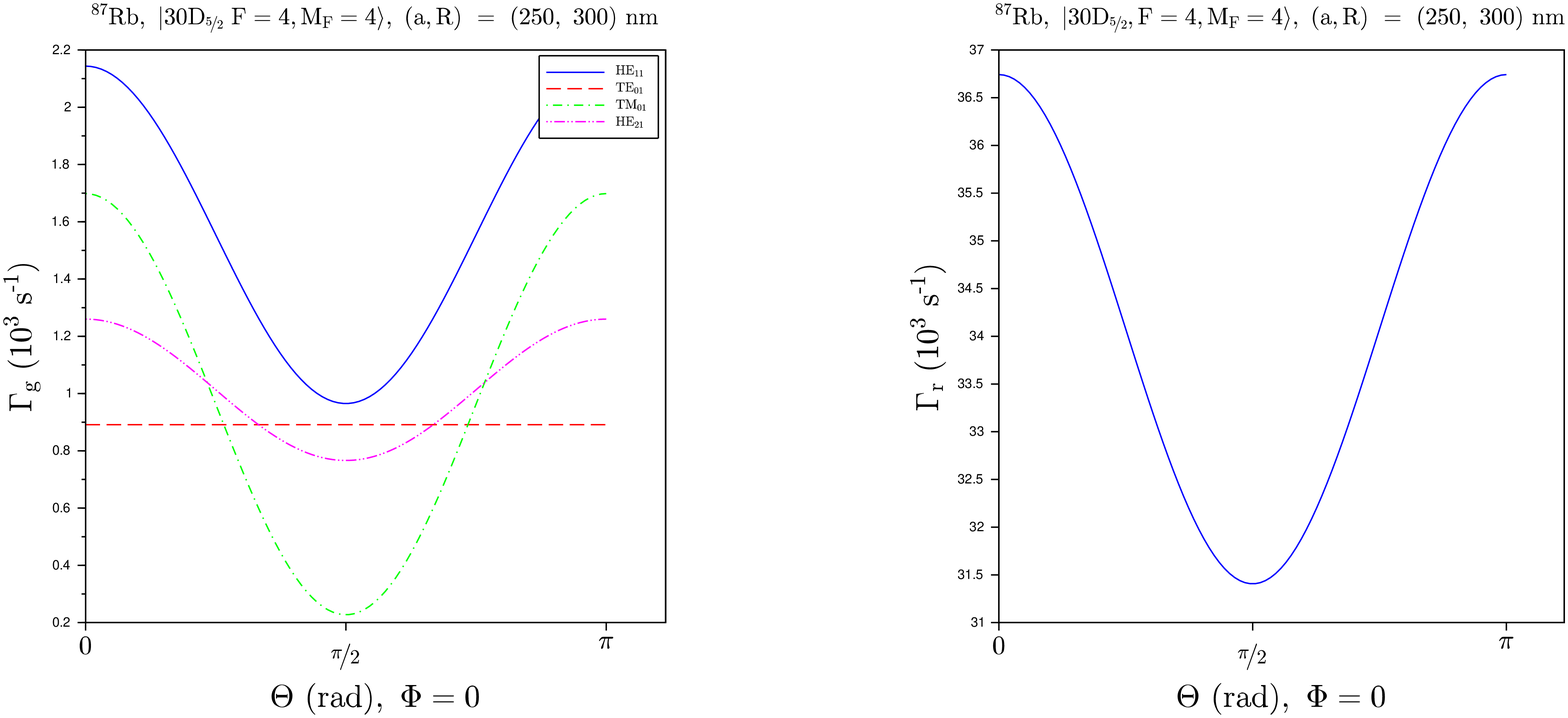} 
\par\end{centering}
\caption{\textbf{Spontaneous emission of an $^{87}\text{Rb}$ atom near an
optical nanofiber with quantization axis in the $\left(Oxz\right)$
plane. }We represent the spontaneous emission rates, $\Gamma_{g}$
(left) and $\Gamma_{r}$ (right), into the first guided and radiative
modes, respectively, for an $^{87}\text{Rb}$ atom in the state $\left|30D_{\nicefrac{5}{2}},F=4,M_{F}=4\right\rangle $
as functions of the angle $\Theta$ (c.f. Fig. \ref{AnglesQuantizationAxis}),
with $\Theta=\nicefrac{\pi}{2}$. The contributions to $\Gamma_{g}$
of the first four guided modes, $\text{HE}_{11},\text{TE}_{01}$,
$\text{\ensuremath{\text{TM}_{01}}}$ and $\text{\ensuremath{\text{HE}_{21}}}$
are displayed separately. The radius of the fiber is $a=250$~nm
and the atom is located $50$~nm from the fiber (i.e., $R=a+50$~nm).}

\label{QuantAxOXZ} 
\end{figure}

\begin{figure}
\begin{centering}
\includegraphics[width=16cm]{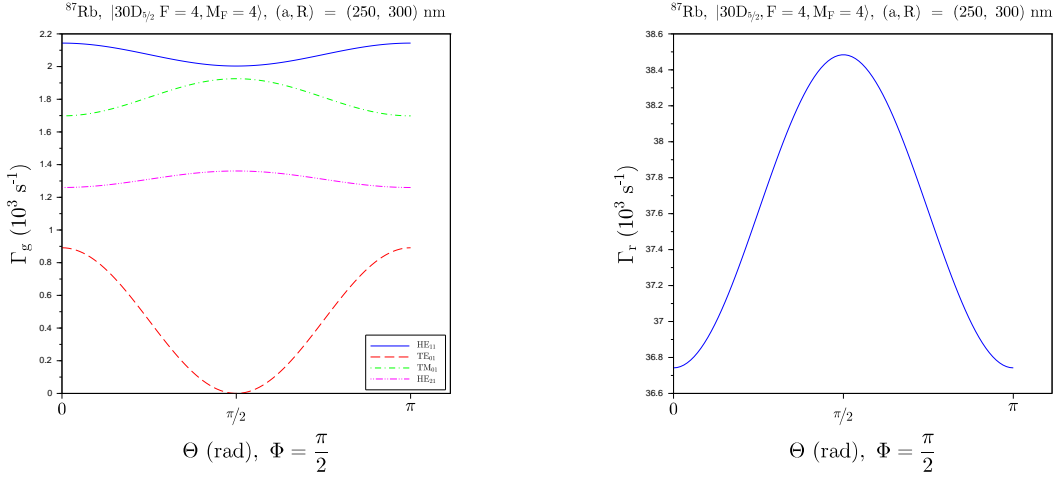} 
\par\end{centering}
\caption{\textbf{Spontaneous emission of an $^{87}\text{Rb}$ atom near an
optical nanofiber with quantization axis in the $\left(Oyz\right)$
plane. }We represent the spontaneous emission rates, $\Gamma_{g}$
(left) and $\Gamma_{r}$ (right), into the first guided and radiative
modes, respectively, for an $^{87}\text{Rb}$ atom in the state $\left|30D_{\nicefrac{5}{2}},F=4,M_{F}=4\right\rangle $
as functions of the angle $\Theta$ (c.f. Fig. \ref{AnglesQuantizationAxis}),
with $\Phi=\nicefrac{\pi}{2}$. The contributions to $\Gamma_{g}$
of the first three guided modes, $\text{HE}_{11},\text{TE}_{01}$
and $\text{\ensuremath{\text{TM}_{01}}}$, are displayed separately.
The radius of the fiber is $a=250$~nm (i.e. $R=a+50$~nm).}

\label{QuantAxOYZ} 
\end{figure}

To be more specific, Figs. \ref{QuantAxOXY}, \ref{QuantAxOXZ} and
\ref{QuantAxOYZ} show the variations of the spontaneous emission
rates towards the first four guided modes, $\Gamma_{g}$ (left), and
towards the radiative modes, $\Gamma_{r}$ (right), for an $^{87}\text{Rb}$
atom prepared in the state $\left|30D_{\nicefrac{5}{2}},F=4,M_{F}=4\right\rangle $
and located at a distance $R=300$~nm from the axis of a silica optical
nanofiber of radius $a=250$~nm when the quantization axis rotates
in the planes $\left(Oxy\right)$, $\left(Oxz\right)$ and $\left(Oyz\right)$,
respectively.

\paragraph{Guided modes}

Before discussing our results on $\Gamma_{g}$ let us make a few remarks
:

A. Owing to our choice of initial atom state, $\left|30D_{\nicefrac{5}{2}},F=4,M_{F}=4\right\rangle $,
and the value of fiber radius considered here, $a=250\text{ nm}$,
the only transitions along which the atom can decay by emitting a
photon into a guided mode are $\sigma^{+}$-transitions towards $P$
states, whose dipole is contained in the plane orthogonal to the quantization
axis.

B. A guided mode is characterized by its type (K=TE, TM, HE, EH),
its frequency $\omega$, two integers $l\geq0$ and $m\geq0$ called
the azimuthal and radial mode orders, respectively, and two numbers
$f=\pm1$ and $p=\pm1$, which characterize the propagation direction
of the mode ($f=\pm1$ conventionally corresponds to a mode propagating
along $\left(Oz\right)$ towards increasing/decreasing $z$) and the
counterclockwise or clockwise phase circulation of the mode, respectively
\citep{KBT17}.

C. Because of field confinement, a guided mode $\mu\equiv\left(\text{K}_{lm},\omega,f,p\right)$
possesses a non-vanishing longitudinal component, $E_{z}^{\left(\mu\right)}$
(except for K=TE) \citep{KR14}. For the guided modes considered,
$E_{z}^{\left(\mu\right)}$ and $E_{y}^{\left(\mu\right)}$can be
chosen as real and $E_{x}^{\left(\mu\right)}$ is then purely imaginary.
Moreover, the mode field components can be written in the form 
\begin{eqnarray*}
E_{x}^{\left(\mu\right)} & = & \text{i}\mathcal{E}_{x}^{\left(\text{K}_{lm},\omega\right)}\\
E_{y}^{\left(\mu\right)} & = & p\mathcal{E}_{y}^{\left(\text{K}_{lm},\omega\right)}\\
E_{z}^{\left(\mu\right)} & = & f\mathcal{E}_{z}^{\left(\text{K}_{lm},\omega\right)}
\end{eqnarray*}
where $\mathcal{E}_{x,y,z}^{\left(\text{K}_{lm},\omega\right)}$ are
real functions of space and time, independent of $f$ and $p$.

D. Finally, note that $\mathcal{E}_{x}^{\left(\text{TE}_{0m},\omega\right)}=\mathcal{E}_{z}^{\left(\text{TE}_{0m},\omega\right)}=0$
and $\mathcal{E}_{y}^{\left(\text{TM}_{0m},\omega\right)}=0$.

Figure \ref{QuantAxOXY} corresponds to the configuration $\Theta\equiv\frac{\pi}{2}$,
i.e., the quantization axis is chosen in the plane $\left(Oxy\right)$
and directed along the vector $\vec{e}_{\text{q}}\equiv\cos\Phi\vec{e}_{x}+\sin\Phi\vec{e}_{y}$.
The dipole, $\vec{d}_{kn}$, associated with the $\sigma^{+}$-de-excitation,
$\left|n\right\rangle \rightarrow\left|k\right\rangle $, of frequency
$\omega_{nk}$, can, therefore, be written in the form $\vec{d}_{kn}=\frac{d_{kn}}{\sqrt{2}}\left[\text{i}\left(\sin\Phi\vec{e}_{x}-\cos\Phi\vec{e}_{y}\right)+\vec{e}_{z}\right]$.
According to the remarks above, the coupling factor $\vec{d}_{kn}\cdot\vec{E}^{\left(\mu\right)}$
of a given transition $\left|n\right\rangle \rightarrow\left|k\right\rangle $
to the (resonant) guided mode $\mu\equiv\left(\text{K}_{lm},\omega_{nk},f,p\right)$
is proportional to $f\mathcal{E}_{z}^{\left(\text{K}_{lm},\omega_{nk}\right)}-\mathcal{E}_{x}^{\left(\text{K}_{lm},\omega_{nk}\right)}\sin\Phi-\text{i}p\mathcal{E}_{y}^{\left(\text{K}_{lm},\omega_{nk}\right)}\cos\Phi$
and the associated contribution to the spontaneous emission rate is,
therefore, itself proportional to $\left(f\mathcal{E}_{z}^{\left(\text{K}_{lm},\omega_{nk}\right)}-\mathcal{E}_{x}^{\left(\text{K}_{lm},\omega_{nk}\right)}\sin\Phi\right)^{2}+\left(\mathcal{E}_{y}^{\left(\text{K}_{lm},\omega_{nk}\right)}\right)^{2}\cos^{2}\Phi$.
Summing over $f=\pm1$, $p=\pm1$ and all possible final states, $\left|k\right\rangle $,
we conclude that the spontaneous emission rate, $\Gamma_{g}^{\left(\text{K}_{lm}\right)}$,
into the first four guided modes $\text{K}_{lm}=\text{HE}_{11}$,
$\text{TE}_{01}$, $\text{TM}_{01}$ and $\text{HE}_{21}$, is proportional
to 
\[
\sum_{k}\left\{ \left(\mathcal{E}_{z}^{\left(\text{K}_{lm},\omega_{nk}\right)}\right)^{2}+\left(\mathcal{E}_{x}^{\left(\text{K}_{lm},\omega_{nk}\right)}\right)^{2}\sin^{2}\Phi+\left(\mathcal{E}_{y}^{\left(\text{K}_{lm},\omega_{nk}\right)}\right)^{2}\cos^{2}\Phi\right\} 
\]
(Note that cross-terms between $E_{z}$ and $E_{x}$ compensate each
other when summing over $f$). In agreement with Fig. \ref{QuantAxOXY},
we conclude that: i) $\Gamma_{g}^{\left(\text{K}_{lm}\right)}$ is
a $\pi$-periodic function of $\Phi$ and reaches its extrema when
$\Phi=0\left[\frac{\pi}{2}\right]$. ii) For the modes $\text{TE}_{01}$,
since $\mathcal{E}_{x}=\mathcal{E}_{z}=0$, $\Gamma_{g}^{\left(\text{TE}_{01}\right)}\left(\Phi\right)\propto\cos^{2}\Phi$
is maximal for $\Phi=0\left[\pi\right]$, minimal for $\Phi=\frac{\pi}{2}\left[\pi\right]$
and its minimum is zero. iii) For the modes $\text{TM}_{01}$, since
$\mathcal{E}_{y}=0$, $\Gamma_{g}^{\left(\text{TM}_{01}\right)}\left(\Phi\right)\propto\sum_{k}\left\{ \left(\mathcal{E}_{z}^{\left(\text{TM}_{01},\omega_{nk}\right)}\right)^{2}+\left(\mathcal{E}_{x}^{\left(\text{TM}_{01},\omega_{nk}\right)}\right)^{2}\sin^{2}\Phi\right\} $
is maximal for $\Phi=\frac{\pi}{2}\left[\pi\right]$, minimal for
$\Phi=0\left[\pi\right]$ and its minimum is different from zero.
For other modes $\left(\text{K}_{lm}=\text{HE}_{11},\text{HE}_{21}\right)$,
Fig. \ref{QuantAxOXY} shows that minima and maxima of $\Gamma_{g}^{\left(\text{K}_{lm}\right)}\left(\Phi\right)$
are also reached for $\Phi=0\left[\pi\right]$ and $\Phi=\frac{\pi}{2}\left[\pi\right]$,
respectively. This can be explained by the inequality $\left|\mathcal{E}_{x}\right|\geq\left|\mathcal{E}_{y}\right|$
valid for these modes and the values $\left(a,R\right)$ considered.

The same arguments can be used to interpret Fig. \ref{QuantAxOXZ}.
This time, the quantization axis is chosen in the plane $\left(Oxz\right)$,
i.e., $\Phi\equiv0$, and $\vec{e}_{\text{q}}\equiv\sin\Theta\vec{e}_{x}+\cos\Theta\vec{e}_{z}$,
whence $\vec{d}_{kn}=\frac{d_{kn}}{\sqrt{2}}\left[\text{i}\left(\sin\Theta\vec{e}_{z}-\cos\Theta\vec{e}_{x}\right)+\vec{e}_{y}\right]$.
The contribution to the spontaneous emission rate into the resonant
guided mode $\mu\equiv\left(\text{K}_{lm},\omega_{nk},f,p\right)$
of a given transition $\left|n\right\rangle \rightarrow\left|k\right\rangle $
is proportional to $\left(\cos\Theta\mathcal{E}_{x}^{\left(\text{K}_{lm},\omega_{nk}\right)}+p\mathcal{E}_{y}^{\left(\text{K}_{lm},\omega_{nk}\right)}\right)^{2}+\left(\mathcal{E}_{z}^{\left(\text{K}_{lm},\omega_{nk}\right)}\right)^{2}\sin^{2}\Theta$.
Summing over $f=\pm1$, $p=\pm1$, and $k$, we conclude that the
spontaneous emission rate $\Gamma_{g}^{\left(\text{K}_{lm}\right)}$
into guided modes $\text{K}_{lm}=\text{HE}_{11}$, $\text{TE}_{01}$,
$\text{TM}_{01}$ or $\text{HE}_{21}$, is proportional to 
\[
\sum_{k<n}\left\{ \cos^{2}\Theta\left(\mathcal{E}_{x}^{\left(\text{K}_{lm},\omega_{nk}\right)}\right)^{2}+\left(\mathcal{E}_{y}^{\left(\text{K}_{lm},\omega_{nk}\right)}\right)^{2}+\left(\mathcal{E}_{z}^{\left(\text{K}_{lm},\omega_{nk}\right)}\right)^{2}\sin^{2}\Theta\right\} 
\]
(Note that cross-terms between $E_{x}$ and $E_{y}$ now compensate
each other when summing over $p$). In agreement with Fig. \ref{QuantAxOXZ},
we conclude that : i) $\Gamma_{g}^{\left(\text{K}_{lm}\right)}$ is
a $\pi$-periodic function of $\Theta$ which reaches its extrema
for $\Theta=0\left[\frac{\pi}{2}\right]$. ii) For the modes $\text{TE}_{01}$,
since $\mathcal{E}_{x}=\mathcal{E}_{z}=0$, $\Gamma_{g}^{\left(\text{TE}_{01}\right)}\left(\Theta\right)$
is constant. iii) For other modes $\left(\text{K}_{lm}=\text{HE}_{11},\text{TM}_{01},\text{HE}_{21}\right)$,
Fig. \ref{QuantAxOXZ} shows that maxima and minima are achieved for
$\Theta=0\left[\pi\right]$ and $\Theta=\frac{\pi}{2}\left[\pi\right]$,
respectively, i.e., $\Gamma_{g,\text{max}}\propto\sum_{\alpha}\left\{ \left(\mathcal{E}_{x}^{\left(\text{K}_{lm},\omega_{nk}\right)}\right)^{2}+\left(\mathcal{E}_{y}^{\left(\text{K}_{lm},\omega_{nk}\right)}\right)^{2}\right\} $
and $\Gamma_{g,\text{min}}\propto\sum_{\alpha}\left\{ \left(\mathcal{E}_{y}^{\left(\text{K}_{lm},\omega_{nk}\right)}\right)^{2}+\left(\mathcal{E}_{z}^{\left(\text{K}_{lm},\omega_{nk}\right)}\right)^{2}\right\} $.
This can be explained by the inequality $\left|\mathcal{E}_{x}\right|\geq\left|\mathcal{E}_{z}\right|$
valid for these modes and the values $\left(a,R\right)$ considered.

Finally, in Fig. \ref{QuantAxOYZ}, the quantization axis is chosen
in the plane $\left(Oyz\right)$, i.e. $\Phi\equiv\nicefrac{\pi}{2}$,
and $\vec{e}_{\text{q}}\equiv\sin\Theta\vec{e}_{y}+\cos\Theta\vec{e}_{z}$,
whence $\vec{d}_{kn}=\frac{d_{kn}}{\sqrt{2}}\left[\text{i}\left(\cos\Theta\vec{e}_{y}-\sin\Theta\vec{e}_{z}\right)+\vec{e}_{x}\right]$.
The contribution to the spontaneous emission rate into the resonant
guided mode $\mu\equiv\left(\text{K}_{lm},\omega_{nk},f,p\right)$
of a given transition $\left|n\right\rangle \rightarrow\left|k\right\rangle $
is proportional to $\left[\mathcal{E}_{x}^{\left(\text{K}_{lm},\omega_{nk}\right)}+\cos\Theta p\mathcal{E}_{y}^{\left(\text{K}_{lm},\omega_{nk}\right)}-\sin\Theta f\mathcal{E}_{z}^{\left(\text{K}_{lm},\omega_{nk}\right)}\right]^{2}$.
Summing over $f=\pm1$, $p=\pm1$, and $k$, we conclude that the
spontaneous emission rate, $\Gamma_{g}^{\left(\text{K}_{lm}\right)}$,
into guided modes of type $\text{K}_{lm}=\text{HE}_{11}$, $\text{TE}_{01}$,
$\text{TM}_{01}$ and $\text{HE}_{21}$ is proportional to 
\[
\sum_{k<n}\left\{ \left(\mathcal{E}_{x}^{\left(\text{K}_{lm},\omega_{nk}\right)}\right)^{2}+\cos^{2}\Theta\left(\mathcal{E}_{y}^{\left(\text{K}_{lm},\omega_{nk}\right)}\right)^{2}+\sin^{2}\Theta\left(\mathcal{E}_{z}^{\left(\text{K}_{lm},\omega_{nk}\right)}\right)^{2}\right\} 
\]
(Note that cross-terms between $E_{x}$, $E_{y}$ and $E_{z}$ now
compensate each other when summing over $p$ and $f$). In agreement
with Fig. \ref{QuantAxOYZ}, we conclude that : i) $\Gamma_{g}^{\left(\text{K}_{lm}\right)}$
is a $\pi$-periodic function of $\Theta$ which reaches its extrema
in $\Theta=0\left[\frac{\pi}{2}\right]$. ii) For the modes $\text{TE}_{01}$,
since $\mathcal{E}_{x}=\mathcal{E}_{z}=0$, $\Gamma_{g}^{\left(\text{TE}_{01}\right)}\left(\Theta\right)\propto\cos^{2}\Theta$
is maximal for $\Theta=0\left[\pi\right]$, minimal for $\Theta=\frac{\pi}{2}\left[\pi\right]$
and its minimum is zero. According to Fig. \ref{QuantAxOYZ}, $\Gamma_{g}^{\left(\text{HE}_{11}\right)}\left(\Theta\right)$
also reaches its maxima and minima in $\Theta=0\left[\pi\right]$
and $\Theta=\frac{\pi}{2}\left[\pi\right]$, respectively. This can
be explained by the inequality $\left|\mathcal{E}_{y}^{\text{HE}_{11}}\right|\geq\left|\mathcal{E}_{z}^{\text{HE}_{11}}\right|$
valid for the values $\left(a,R\right)$ considered. iii) For the
modes $\text{TM}_{01}$, since $\mathcal{E}_{y}=0$, $\Gamma_{g}^{\left(\text{TM}_{01}\right)}\left(\Theta\right)\propto\sum_{k<n}\left\{ \left(\mathcal{E}_{x}^{\left(\text{TM}_{01},\omega_{nk}\right)}\right)^{2}+\sin^{2}\Theta\left(\mathcal{E}_{z}^{\left(\text{TM}_{01},\omega_{nk}\right)}\right)^{2}\right\} $
is maximal for $\Theta=\frac{\pi}{2}\left[\pi\right]$, minimal for
$\Theta=0\left[\pi\right]$. According to Fig. \ref{QuantAxOYZ},
$\Gamma_{g}^{\left(\text{HE}_{21}\right)}\left(\Theta\right)$ also
reaches its maxima and minima in $\Theta=\frac{\pi}{2}\left[\pi\right]$
and $\Theta=0\left[\pi\right]$, respectively. This can be explained
by the inequality $\left|\mathcal{E}_{z}\right|\geq\left|\mathcal{E}_{y}\right|$
valid for $\text{HE}_{21}$ modes and the values $\left(a,R\right)$
considered.

\paragraph{Radiative modes}

Our results on the spontaneous emission rate into the radiative modes
are displayed in the right-hand panels of Figs. \ref{QuantAxOXY},
\ref{QuantAxOXZ} and \ref{QuantAxOYZ}. In the three different configurations,
one observes a $\pi$-periodicity in $\left(\Phi,\Theta\right)$.
Moreover, the three figures seem to indicate that, for the values
of $\left(a,R\right)$ considered, radiative modes contributing to
$\Gamma_{r}$ are mainly radial, i.e., their component along $\left(Ox\right)$
dominates. Due to the variety and complexity of the structure of radiative
modes, it is, however, difficult to go further into the interpretation
of our results.

\paragraph{Proportion of spontaneously emitted light towards the guided modes}

\begin{figure}
\begin{centering}
\includegraphics[width=16cm]{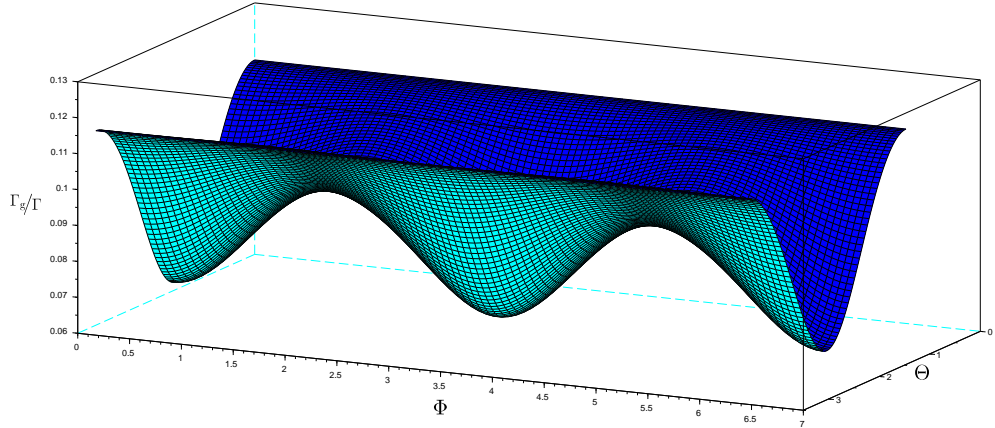} 
\par\end{centering}
\caption{\textbf{Spontaneous emission of an $^{87}\text{Rb}$ atom near an
optical nanofiber.} We represent the proportion of spontaneous emission
into the guided modes, $\nicefrac{\Gamma_{g}}{\Gamma}$, for an $^{87}\text{Rb}$
atom in the state $|\mathrm{30D}_{5/2},F=4,M_{F}|=4\rangle$ as a
function of the angles $\left(\Theta,\Phi\right)$ (c.f. Fig. \ref{AnglesQuantizationAxis}).}

\label{QuantAx3D} 
\end{figure}

Figure \ref{QuantAx3D} displays a 3D ``summary'' of Figs. \ref{QuantAxOXY},
\ref{QuantAxOXZ} and \ref{QuantAxOYZ}. To be more explicit, it shows
the ratio $\nicefrac{\Gamma_{g}}{\Gamma}$ characterizing the proportion
of spontaneous emitted light captured by guided modes. Note that the
contribution of $\text{HE}_{11}$ to $\Gamma_{g}$ dominates. Besides
$\pi$-periodicity in $\Phi$ and $\Theta$, one observes maxima for
$\nicefrac{\Gamma_{g}}{\Gamma}$ for $\vec{e}_{\text{q}}=\vec{e}_{z}$
and saddle points for $\vec{e}_{\text{q}}=\vec{e}_{y}$.

\subsubsection{Anisotropic spontaneous emission\label{GamAnis}}

Througout this section, the quantization axis is chosen along $\left(Oy\right)$.
Using the same notations as in the previous section, this corresponds
to $\vec{e}_{\text{q}}=\vec{e}_{y}$. In this configuration, the atomic
dipole associated with, e.g., a $\sigma^{+}$-transition $\left|n\right\rangle \rightarrow\left|k\right\rangle $
lies in the plane $\left(Oxz\right)$ and, more explicitly, $\vec{d}_{kn}=\frac{d_{kn}}{\sqrt{2}}\left[\text{i}\vec{e}_{x}+\vec{e}_{z}\right]$.
Using, as in the previous section, the simplistic mode function approach,
we conclude that the contribution of this transition to the spontaneous
emission rate into a specific guided mode $\mu=\left(\text{K}_{lm},\omega_{nk},f,p\right)$
is proportional to $\left(f\mathcal{E}_{z}^{\left(K_{lm},\omega_{nk}\right)}-\mathcal{E}_{x}^{\left(K_{lm},\omega_{nk}\right)}\right)^{2}$
and clearly depends on the propagation direction, $f$. This heuristic
argument cannot be straightforwardly transposed to radiative modes,
but the same phenomenon is observed. The anisotropic spontaneous emission
leads to a non-vanishing average lateral force on the atom whose order
of magnitude is $0.5$~zN (5~zN) for a rubidium atom in a $5D$
$\left(5P\right)$ state located at a distance $d=50$~nm from a
fiber of radius $a=200$~nm. This force corresponds to the resonant
part of the average Lorentz force, $\left[F^{\text{res}}\right]_{z}$,
Eq. (\ref{EqFRes}) \citep{Buh12}, and can be calculated in the Green's
function approach. In particular, for an atom initially in a state
$\left|n\right\rangle $, one can decompose $\left[F^{\text{res}}\right]_{z}$
as the sum of contributions $\left[F_{nk,\nu}^{\text{res}}\right]_{z}$
relative to the transition $\left|n\right\rangle \rightarrow\left|k\right\rangle $
coupled to the (guided or radiative) mode, $\nu$.

In order to quantitatively characterize the anisotropy of emission,
we introduce the factor 
\[
\alpha_{n}\equiv\sum_{\nu,k<n}\frac{\Gamma_{nk,\nu}}{\Gamma_{n}}\times\frac{\hbar k_{\nu,z}}{\hbar k_{\nu}}
\]
where the sum runs over all (radiative and guided) modes, $\nu$,
and final states, $k$. In this expression, $\Gamma_{nk,\nu}$ represents
the spontaneous emission rate for the transition $\left|n\right\rangle \rightarrow\left|k\right\rangle $
into the mode $\nu$, $\Gamma_{n}$ is the total spontaneous emission
rate from the state $\left|n\right\rangle $, $k_{\nu,z}$ is the
projection onto $\left(Oz\right)$ of the wavevector for the (guided
or radiative) mode $\left(\nu\right)$ and $k_{\nu}=\nicefrac{\omega_{\nu}}{c}$
is its norm. With these definitions, $\left(\nicefrac{\Gamma_{nk,\nu}}{\Gamma_{n}}\right)$
can be interpreted as the probability for a photon to be emitted from
the state $\left|n\right\rangle $ via the transition $\left|n\right\rangle \rightarrow\left|k\right\rangle $
and into the mode $\nu$, while $\nicefrac{\hbar k_{\nu,z}}{\hbar k_{\nu}}$
characterizes the inclination of the momentum of the photon emitted
into the mode $\nu$ with respect to the fiber axis.

\begin{figure}
\centering{}\includegraphics[width=16cm]{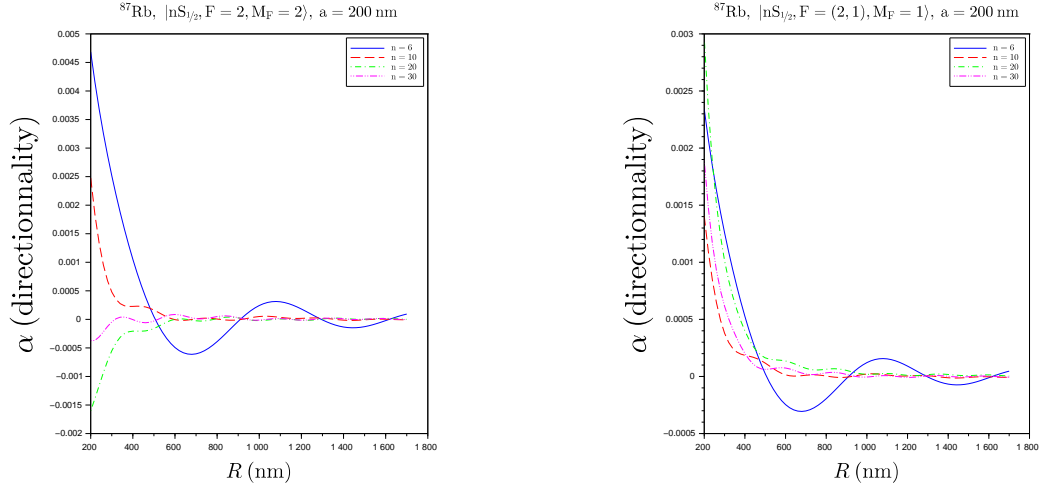}\caption{\textbf{Directionality of the spontaneous emission of an $^{87}\text{Rb}$
atom near an optical nanofiber.} We represent the coefficient, $\alpha_{n}$
(see main text for definition), characterizing the directionality
of the spontaneous emission with respect to the $z$-axis, of an $^{87}\text{Rb}$
atom in the states $|\mathrm{nS}_{1/2},F=2,M_{F}=2\rangle$ (left
panel) and $|\mathrm{nS}_{1/2},F=\left(2,1\right),M_{F}=1\rangle$
(right panel) for $n=6,10,20,30$, close to an optical nanofiber of
radius $a=200$~nm as a function of the distance, $R$, from the
atom to the fiber axis. }
\label{FigAnisotropyS}
\end{figure}

\begin{figure}
\begin{centering}
\includegraphics[width=16cm]{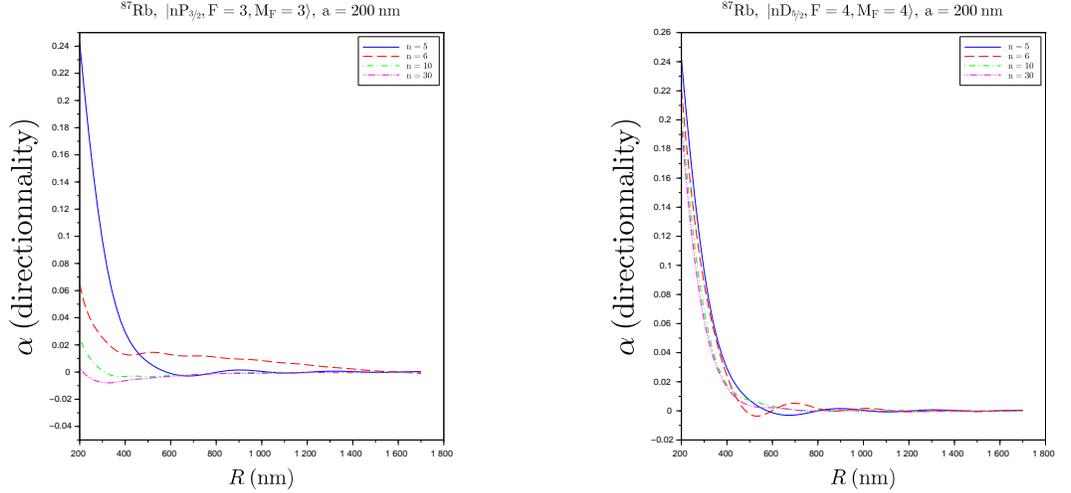} 
\par\end{centering}
\caption{\textbf{Directionality of the spontaneous emission of an $^{87}\text{Rb}$
atom near an optical nanofiber --} We represent the coefficient,
$\alpha_{n}$ (see main text for definition), characterizing the directionality
with respect to the $z$ axis, of the spontaneous emission of an $^{87}\text{Rb}$
atom in the states $|\mathrm{nP}_{3/2},F=3,M_{F}=3\rangle$ (left
panel) and $|\mathrm{nD}_{5/2},F=4,M_{F}=4\rangle$ (right panel)
for $n=5,6,10,30$, close to an optical nanofiber of radius $a=200$~nm
as a function of the distance, $R$, from the atom to the fiber axis. }

\label{FigAnisotropyPD} 
\end{figure}

Identifying $-\hbar k_{\nu,z}\Gamma_{nk,\nu}$, i.e., the atomic recoil
along $\left(Oz\right)$ induced by the emission of a photon into
the mode, $\nu$, via the transition $\left|n\right\rangle \rightarrow\left|k\right\rangle $,
with the force $\left[F_{nk,\nu}^{\text{res}}\right]_{z}$, one can
write $\alpha_{n}=-\sum_{k,\nu}\nicefrac{\left[F_{nk,\nu}^{\text{res}}\right]_{z}}{\Gamma_{n}\hbar k_{\nu}}$
(see \citep{SBC15} and Appendix \ref{AppF}). Figs. \ref{FigAnisotropyS}
and \ref{FigAnisotropyPD} show the coefficient $\alpha_{n}$ for
an $^{87}\text{Rb}$ atom prepared in an excited $S,P$ or $D$ state
decaying via $\sigma^{+}$-transitions located close to an optical
nanofiber of radius $a=200$~nm as a function of the distance $R$
from the atom to the fiber axis, $Oz$. The observed Drexhage-like
oscillations are due to radiative modes \citep{D70}. Remarkably,
though very weak, the spontaneous emission anisotropy for the $S$
states is nonzero, at around $0.4\%$ at most (see Fig. \ref{FigAnisotropyS}).
For $S$ states, $\alpha$ decreases for increasing $n$ and vanishes
when $R\rightarrow+\infty$ as expected (equivalent to the free-space
configuration). As seen in Fig. \ref{FigAnisotropyPD}, for $P$ and
$D$ states, the spontaneous emission anisotropy, at around $20\%$
on the surface of the nanofiber, is much stronger than for $S$ states.
When $R\rightarrow+\infty$, $\alpha_{n}$ tends to zero as expected.
For $P$ states, $\alpha_{n}$ decreases with $n$, while it only
slightly varies for $D$ states. Anisotropic emission is, therefore,
observable for $D$ states even at high values of $n$.

\paragraph{Anisotropic spontaneous emission into the guided modes of the nanofiber}

\begin{figure}
\begin{centering}
\includegraphics[width=16cm]{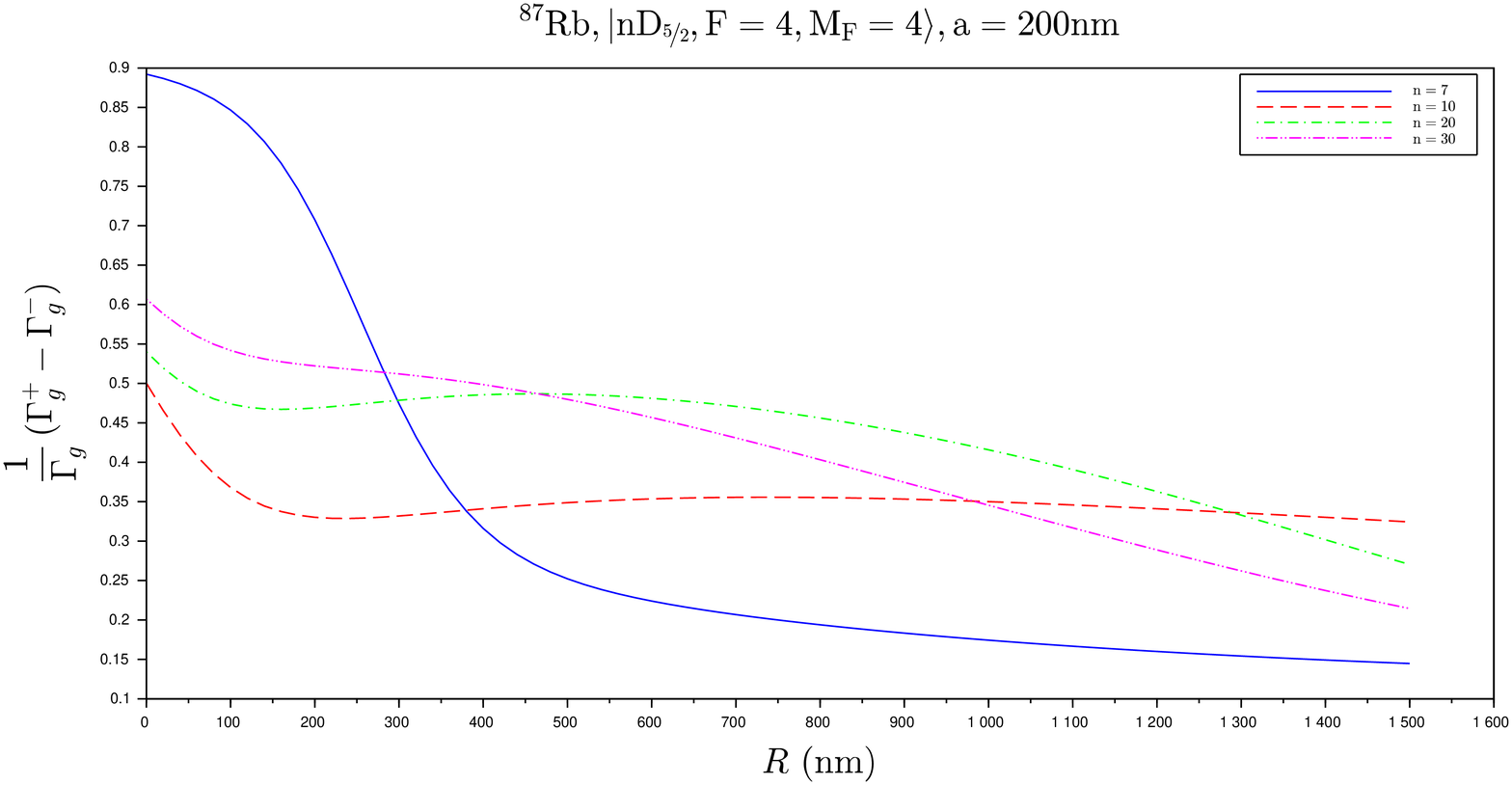} 
\par\end{centering}
\caption{\textbf{Directionality of the spontaneous emission of an $^{87}\text{Rb}$
atom near an optical nanofiber into the guided modes --} We represent
the ratio, $\nicefrac{\left(\Gamma_{g}^{\left(+\right)}-\Gamma_{g}^{\left(-\right)}\right)}{\Gamma_{g}}$
(see main text for definitions), characterizing the directionality
with respect to the $z$ axis, of the spontaneous emission into the
guided modes of an $^{87}\text{Rb}$ atom in the state $|\mathrm{nD}_{5/2},F=4,M_{F}=4\rangle$
for $n=7,10,20,30$, close to an optical nanofiber of radius $a=200$~nm
as a function of the distance, $R$, from the atom to the fiber axis. }

\label{AnisotropyGuided} 
\end{figure}

For guided modes, the anisotropy can be further characterized by the
ratio, $\nicefrac{\left(\Gamma_{g}^{\left(+\right)}-\Gamma_{g}^{\left(-\right)}\right)}{\Gamma_{g}}$,
where $\Gamma_{g}^{\left(\pm\right)}$ denotes the spontaneous emission
rate into forward/backward propagating guided modes and $\Gamma_{g}\equiv\Gamma_{g}^{+}+\Gamma_{g}^{-}$.
Using the same arguments as above, one can write this factor in the
following form: $-\sum_{k,\mu}\nicefrac{\left[F_{nk,\mu}^{\text{res}}\right]_{z}}{\Gamma_{g}\hbar\left|k_{\mu,z}\right|}$,
where now the sum runs over the guided modes, $\mu$, only (see \citep{KR14}
and Appendix \ref{AppF}). Figure \ref{AnisotropyGuided} shows the
ratio $\nicefrac{\left(\Gamma_{g}^{\left(+\right)}-\Gamma_{g}^{\left(-\right)}\right)}{\Gamma_{g}}$
calculated for an $^{87}\text{Rb}$ atom prepared in the state $|\mathrm{nD}_{5/2},F=4,M_{F}=4\rangle$,
with $n=7,10,20,30$, and located near an optical nanofiber of radius
$a=200$~nm, as a function of the distance, $R$, from the atom to
the fiber axis. The directionality of the guided emitted light remains
strong even for high values of $n$ and $R$. Note, however, that
for large $R>300$~nm the absolute value of $\Gamma_{g}$ itself
is so small that the directionality has little practical meaning.

\subsection{Lamb shift and van der Waals force}

\begin{figure}
\noindent \begin{raggedright}
\includegraphics[width=16cm]{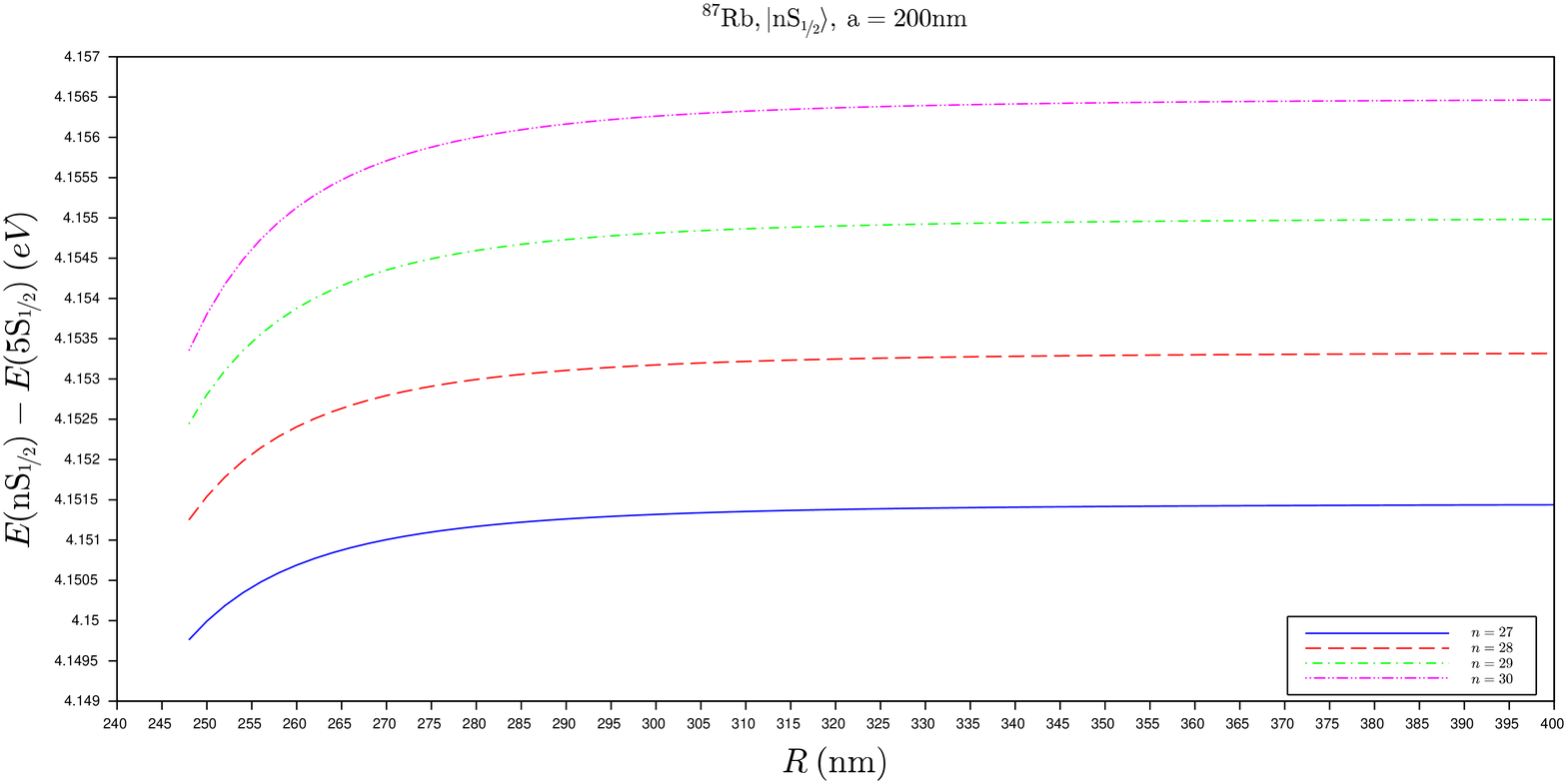} 
\par\end{raggedright}
\caption{\textbf{Lamb shift of an $^{87}\text{Rb}$ atom in the state $\left|nS_{\nicefrac{1}{2}}\right\rangle $,
for $n=27,\cdots,30$ near an optical nanofiber --} We represent
the energy difference, $E\left(nS_{\nicefrac{1}{2}}\right)-E\left(5S_{\nicefrac{1}{2}}\right)$,
of the states $\left|nS_{\nicefrac{1}{2}}\right\rangle $ ($n=27\cdots30$)
and $\left|5S_{\nicefrac{1}{2}}\right\rangle $ of an $^{87}\text{Rb}$
atom near an optical nanofiber of radius, $a=200$~nm as a function
of the distance, $R$, from the fiber. Energies are given in eV. }
\label{LambShiftS} 
\end{figure}

Figure \ref{LambShiftS} displays the energy difference, $E\left(nS_{\nicefrac{1}{2}}\right)-E\left(5S_{\nicefrac{1}{2}}\right)$,
of the states $\left|nS_{\nicefrac{1}{2}}\right\rangle $ ($n=27\cdots30$)
and $\left|5S_{\nicefrac{1}{2}}\right\rangle $ for an $^{87}\text{Rb}$
atom near an optical nanofiber of radius $a=200$~nm as a function
of the distance, $R$, from the fiber axis. The Lamb shift of the
ground state is assumed to be negligible with respect to that of the
excited levels. When $R$ decreases, $\left[E\left(nS_{\nicefrac{1}{2}}\right)-E\left(5S_{\nicefrac{1}{2}}\right)\right]$
itself decreases, though more rapidly for higher $n$. At shorter
distances from the fiber, energy curves cross (not shown on Fig. \ref{LambShiftS})
and the perturbative approach fails. The treatment of this area requires
the diagonalization of the full Hamiltonian in the relevant degenerate
Hilbert subspace. This will be investigated in future work.

\begin{figure}
\begin{centering}
\includegraphics[width=16cm]{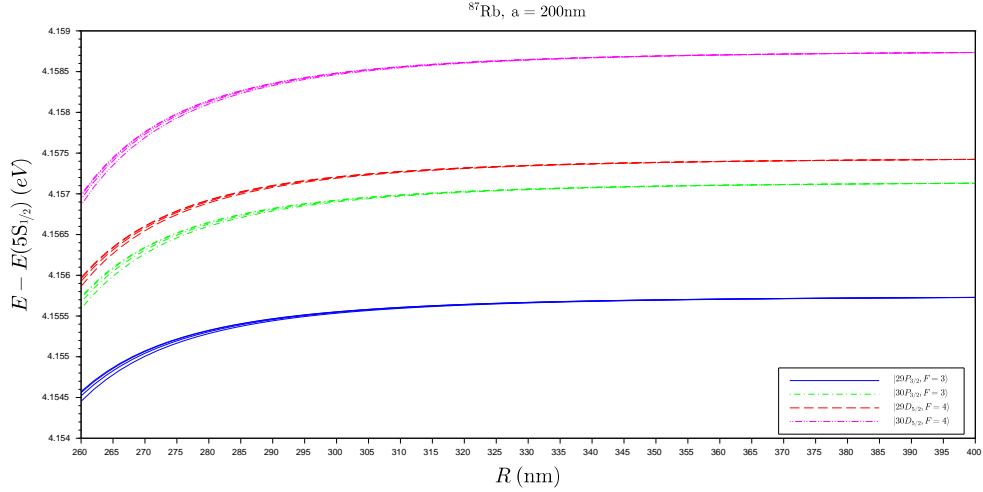} 
\par\end{centering}
\caption{\textbf{Lamb shift of an $^{87}\text{Rb}$ atom in the states $\left|nP_{\nicefrac{3}{2}}F=3,M_{F}=-F\cdots F\right\rangle $
and $\left|nD_{\nicefrac{5}{2}}F=4,M_{F}=-F\cdots F\right\rangle $,
for $n=29,30$ near an optical nanofiber --} We represent the energy
difference, $E-E\left(5S_{\nicefrac{1}{2}}\right)$, of the states
of interest with respect to $\left|5S_{\nicefrac{1}{2}}\right\rangle $
as a function of the distance, $R$, from the fiber. The radius of
the nanofiber is $a=200$~nm. Energies are given in eV. }
\label{LambShiftPD} 
\end{figure}

Figure \ref{LambShiftPD} shows the same quantity for states $\left|nD_{\nicefrac{5}{2}}F=4,m_{F}=-F\cdots F\right\rangle $
and 

$\left|nP_{\nicefrac{3}{2}}F=3,m_{F}=-F\cdots F\right\rangle $ for
$n=29,30$. Though the order of magnitude is comparable to that obtained
for states $\left|nS_{\nicefrac{1}{2}}\right\rangle $, one observes
a degeneracy lift of the hyperfine components of different $\left|M_{F}\right|$
very close to the fiber; to be more explicit, the Lamb shift is stronger
for states of higher $\left|M_{F}\right|$. This can be qualitatively
justified as follows: i) Radiative and guided modes have a strong
-- though not exclusive -- transverse component, i.e., orthogonal
to the fiber axis $\left(Oz\right)$ (see Fig. \ref{System}); ii)
High coupling to the guided modes is, therefore, obtained for transitions
corresponding to dipoles in the transverse plane, $\left(Oxy\right)$;
iii) The quantization axis being along the fiber axis, dipoles in
the plane $\left(Oxy\right)$ correspond to $\sigma$ transitions:
therefore, the stronger the weight of $\sigma$ transitions in the
de-excitation of an excited state, the higher the spontaneous emission
rate into guided modes; iv) The higher $\left|M_{F}\right|$, the
stronger the weight of $\sigma$ transitions in the de-excitation
of the state (this can be directly checked on $3j$-coefficients):
therefore, the higher $\left|M_{F}\right|$, the higher the spontaneous
emission rate into guided modes.

\begin{figure}
\begin{centering}
\includegraphics[width=16cm]{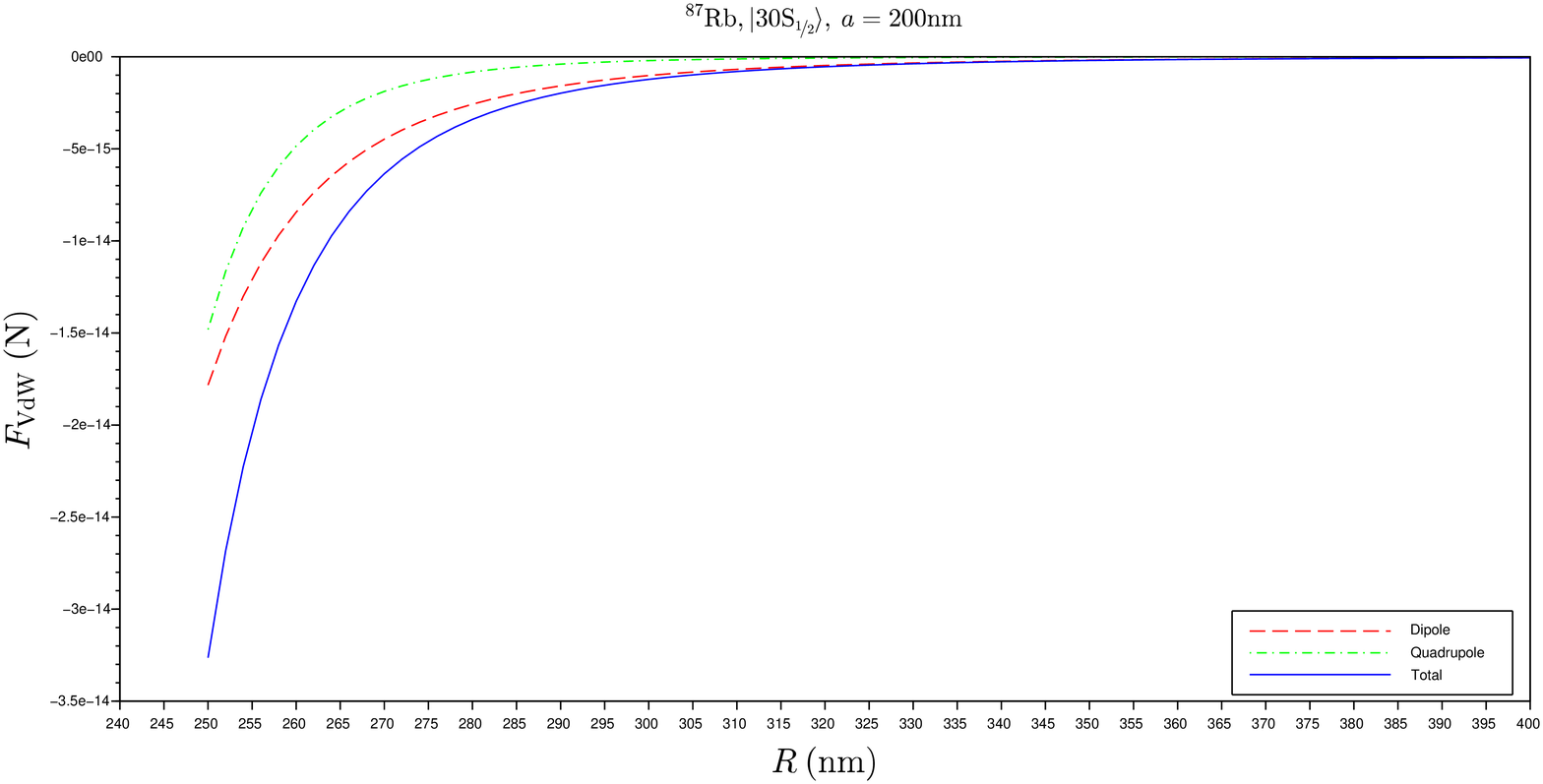} 
\par\end{centering}
\caption{\textbf{van der Waals force felt by an $^{87}\text{Rb}$ atom in the
state $\left|30S_{\nicefrac{1}{2}}\right\rangle $, for $n=25,\cdots,30$
near an optical nanofiber --} We represent the radial van der Waals
force, $F_{vdW}=-\partial_{R}U\left(R\right)$, felt by an $^{87}\text{Rb}$
atom in the state $\left|30S_{\nicefrac{1}{2}}\right\rangle $ close
to an optical nanofiber of radius $a=200$~nm as a function of the
distance, $R$, from the fiber. The total force, electric dipole,
and quadrupole coupling contributions are represented by (blue) full,
(red) dashed, and (green) dashed-dotted lines, respectively. }
\label{FvdWS} 
\end{figure}

The $R$-dependence of the Lamb shift results in a radial van der
Waals force, $-\partial_{R}U_{n}\left(R\right)$, represented in Fig.
\ref{FvdWS} for the state $\left|30S_{\nicefrac{1}{2}}\right\rangle $
as a function of $R$. Note the negative sign and, therefore, the
attractive character of the force, as well as its order of magnitude
of $10^{-14}\text{N}$, much larger than spontaneous emission recoil
induced forces. Aside from the total force, we represented the contributions
of the electric dipole and quadrupole couplings. Though the dipole
contribution dominates, the quadrupolar component is far from negligible,
especially close to the nanofiber when field inhomogeneities are magnified.

\begin{figure}
\begin{centering}
\includegraphics[width=16cm]{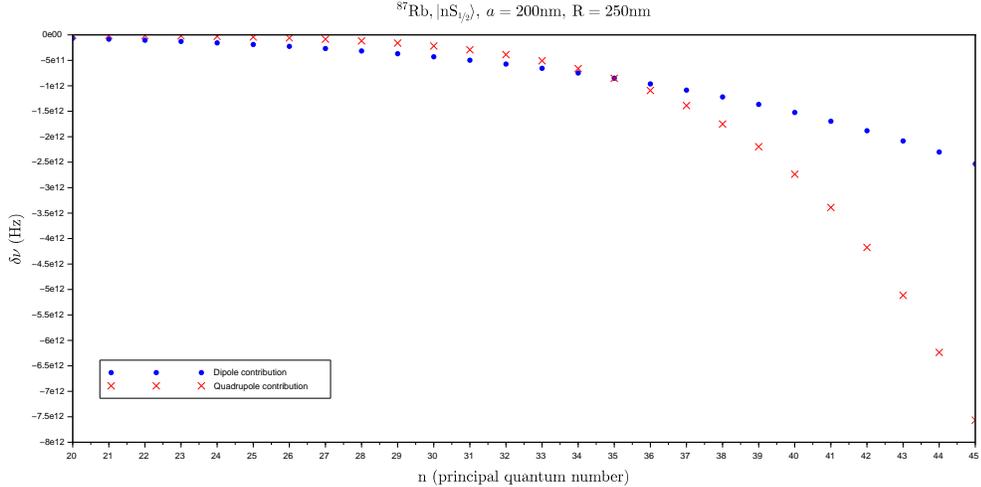} 
\par\end{centering}
\caption{\textbf{Electric dipole and quadrupole contributions to the Lamb shift
of an $^{87}\text{Rb}$ atom in the state $\left|nS_{\nicefrac{1}{2}}\right\rangle $
near an optical nanofiber --} Electric dipole and quadrupole components
are represented as functions of the principal quantum number, $n$,
by (blue) dots and (red) crosses, respectively. The radius of the
optical nanofiber is $a=200$~nm and the atom is located at $R=250$~nm
from the fiber axis.}

\label{LambShiftQuadruN} 
\end{figure}

Figure \ref{LambShiftQuadruN} displays the electric dipole and quadrupole
components of the Lamb shift calculated for an $^{87}\text{Rb}$ atom
in the state $\left|nS_{\nicefrac{1}{2}}\right\rangle $ located at
a distance, $R=250$~nm from an optical nanofiber of radius $a=200$~nm.
One observes that the higher the principal quantum number, $n$, the
stronger the quadrupole component. For $n>35$, it even dominates
the Lamb shift.

\begin{figure}
\begin{centering}
\includegraphics[angle=90,height=20cm]{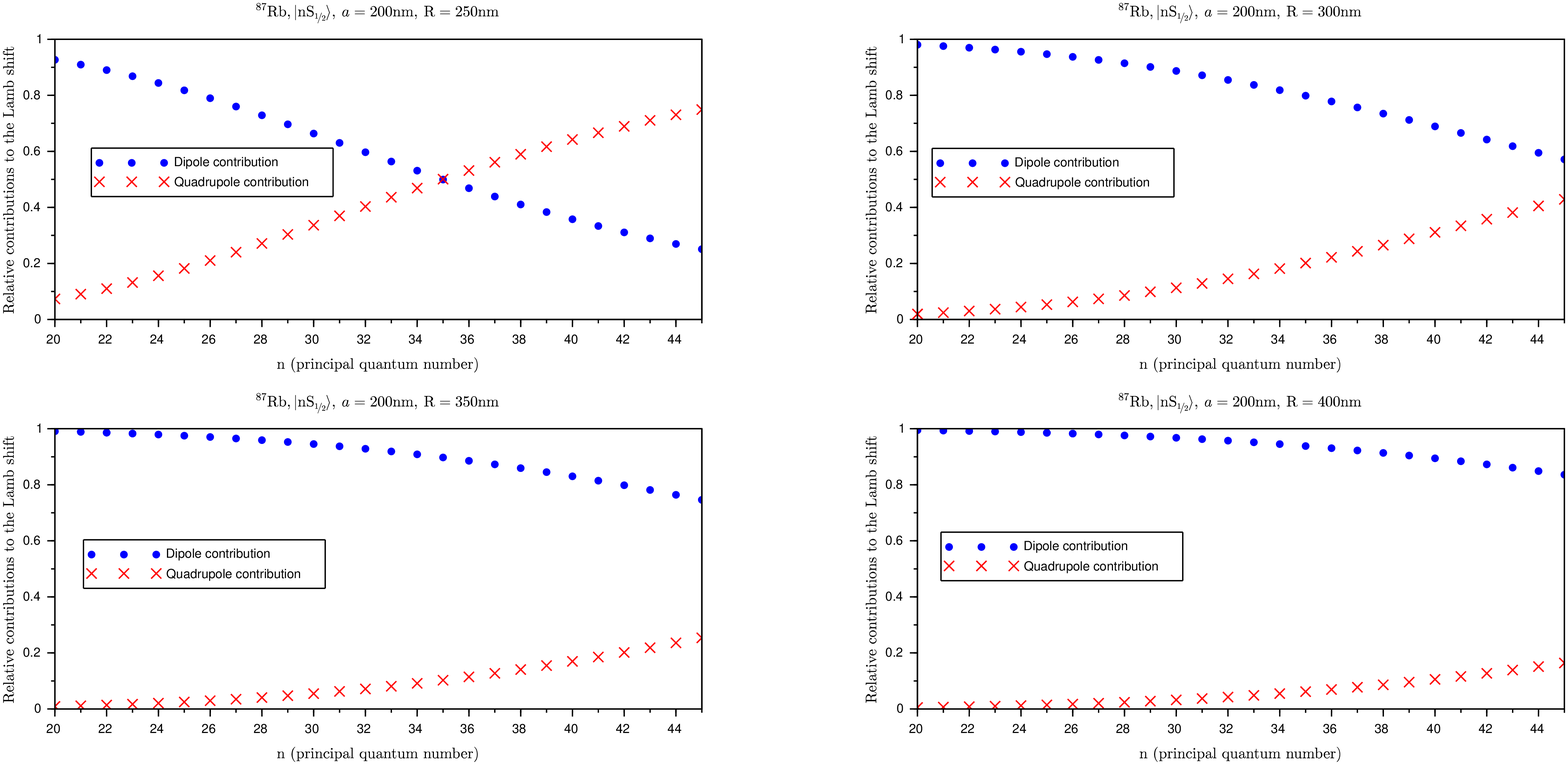} 
\par\end{centering}
\caption{\textbf{Relative contributions to the Lamb shift of electric dipole
and quadrupole couplings for an $^{87}\text{Rb}$ atom in the state
$\left|nS_{\nicefrac{1}{2}}\right\rangle $ close to an optical nanofiber
--} Electric dipole and quadrupole components are represented as
functions of the principal quantum number, $n$, by (blue) dots and
(red) crosses, respectively, for an atom located at $R=250\text{ (top left)},300\text{ (top right)},350\text{ (bottom left)}$
and $400$~nm $\text{ (bottom right)}$ from the fiber axis. The
radius of the optical nanofiber is $a=200$~nm.}

\label{LambShiftQuadruNR} 
\end{figure}

One observes the same trend with $n$ in Fig. \ref{LambShiftQuadruNR},
which displays the relative contributions of the electric dipole and
quadrupole couplings to the Lamb shift calculated for an $^{87}\text{Rb}$
atom in the state $\left|nS_{\nicefrac{1}{2}}\right\rangle $\textbf{
}located at four different distances $R=250,300,350,$ and $400$~nm
from the optical nanofiber axis, as functions of $n$. As expected,
the influence of quadrupolar transitions is lowered when the distance,
$R$, increases, since the effect of the fiber on the electromagnetic
field is less pronounced.

\section{Conclusion \label{SECConclusion}}

The influence of a nanofiber near an $^{87}\text{Rb}$ atom prepared
in a Rydberg-excited state, 

$\left|n\leq30;L=S,P,D;JFM_{F}\right\rangle $, on the spontaneous
emission rates and Lamb shift was investigated numerically in detail.
In particular, the dependence of the spontaneous emission rates on
the fiber radius, the distance of the atom to the fiber, the principal
quantum number, $n$, orbital momentum, fine and hyperfine structures
of the state considered, and the direction of angular momentum polarization
were addressed. Close to the nanofiber, a non-negligible fraction
of the emitted light can be captured by guided modes. This fraction
is higher for larger $\left|M_{F}\right|$ but saturates for high
$n$. When the quantum and fiber axes do not coincide, spontaneous
emission into guided modes becomes strongly directional. This directionality
persists even for high $n$. The contribution of quadrupolar transitions
was shown to be negligible for spontaneous emission rates, while they
may dominate Lamb shifts and van der Waals associated forces for high
$n$. Our calculations were performed in the multimode fiber case,
including all atomic transitions, using the general framework of macroscopic
quantum electrodynamics and this allowed us to account for the dispersive
and absorptive characteristics of silica.

Our work is a preliminary step towards the building of a Rydberg-atom-optical-nanofiber
platform. In particular, the collection and guidance of a substantial
part of the spontaneous emitted light along the nanofiber suggests
the possibility of constructing a network of Rydberg atomic ensembles
in the same spirit as described in \citep{BCA12}. The strong directionality
of spontaneous emission observed for specific Rydberg states and quantization
axis is also very promising in view of potential applications in chiral
quantum information protocols \citep{LMS17}. In future works, we
will address the case of several Rydberg atoms in the neighbourhood
of an optical nanofiber. In particular, we shall be interested in
studying how the nanofiber modifies the Rydberg blockade phenomenon
and whether the geometric arrangement of atoms can be used to enhance
the coupling to guided modes.
\begin{acknowledgments}
This research was supported by the Centre National de la Recherche
Scientifique (CNRS) via the grant ``PICS QuaNet''. SNC acknowledges
support from OIST Graduate University and JSPS Grant-in-Aid for Scientific
Research (C) Grant Number 19K05316. The authors thank Antoine Browaeys,
Tridib Ray and Fam Le Kien for fruitful discussions. 
\end{acknowledgments}

\appendix

\section{Dyadic Green's function for a cylindrical nanofiber\label{AppGDF}}

The dyadic Green's function $\overline{\overline{G}}$ used throughout
the main text is the solution of the Helmholtz equation

\begin{equation}
\left[\vec{\nabla}_{\vec{r}}\times\vec{\nabla}_{\vec{r}}\times-\varepsilon\left(\vec{r},\omega\right)\frac{\omega^{2}}{c^{2}}\right]\overline{\overline{G}}\left(\vec{r},\vec{r}',\omega\right)=\delta\left(\vec{r}-\vec{r}'\right)\overline{\overline{I}}\label{HelmholtzApp}
\end{equation}
where the operator, $\vec{\nabla}_{\vec{r}}$, acts on the position
vector, $\vec{r}$, $\overline{\overline{I}}$ is the unit dyadic,
and $\varepsilon=\varepsilon_{1}\left(\omega\right)$ (silica relative
electric permittivity) inside the nanofiber and $\varepsilon=1$ outside.
As shown in \citep{Tai94}, $\overline{\overline{G}}$ splits into
a vacuum term, $\overline{\overline{G}}_{0}$, which is the solution
of Eq. (\ref{HelmholtzApp}) with $\varepsilon\equiv1$ in all space,
and a scattering term, $\overline{\overline{G}}_{\text{sc}}$, due
to the presence of the nanofiber, i.e.,

\[
\overline{\overline{G}}=\overline{\overline{G}}_{0}+\overline{\overline{G}}_{\mathrm{sc}}.
\]
The scattering term, $\overline{\overline{G}}_{\mathrm{sc}}$, can
be decomposed as follows 
\begin{equation}
\overline{\overline{G}}_{\mathrm{sc}}\left(\vec{r},\vec{r}',\omega\right)=\frac{1}{8\pi}\intop_{-\infty}^{+\infty}\mathrm{d}\beta\sum_{n=-\infty}^{+\infty}\overline{\overline{g}}_{n}\left(\tilde{\rho},\tilde{\rho}',\omega,\beta\right)\mathrm{e}^{\mathrm{i}n\left(\phi-\phi'\right)}\mathrm{e}^{\mathrm{i}\beta\left(z-z'\right)}.\label{Gsc}
\end{equation}
where we introduced the cylindrical coordinates $\left(\rho,\phi,z\right)$
and $\left(\rho',\phi',z'\right)$ of the vectors $\vec{r}$ and $\vec{r}'$,
respectively, $\tilde{\rho}\equiv\eta_{2}\rho$, $\tilde{\rho}'\equiv\eta_{2}\rho'$,
$\eta_{j=1,2}\left(\beta\right)\equiv\sqrt{k_{j}\left(\omega\right)^{2}-\beta^{2}}$
and $k_{j=1,2}\left(\omega\right)\equiv\frac{\omega}{c}\sqrt{\varepsilon_{j}\left(\omega\right)}$.
In the cylindrical bases $\left(\vec{e}_{\rho},\vec{e}_{\phi},\vec{e}_{z}\right)$
and $\left(\vec{e}_{\rho'},\vec{e}_{\phi'},\vec{e}_{z}\right)$ associated
to $\vec{r}$ and $\vec{r}'$, defined by $\vec{r}=\rho\vec{e}_{\rho}+z\vec{e}_{z}$
and $\vec{r}=\rho'\vec{e}_{\rho'}+z'\vec{e}_{z}$, respectively (see
Fig. \ref{System}), the components of the dyadic function, $\overline{\overline{g}}_{n}\left(\tilde{\rho},\tilde{\rho}',\omega,\beta\right)$,
take the forms

\begin{eqnarray*}
\left[\overline{\overline{g}}_{n}\left(\tilde{\rho},\tilde{\rho}',\omega,\beta\right)\right]_{\rho\rho'} & = & \mathrm{i}\left[r_{MM}\frac{nH_{n}^{\left(1\right)}\left(\tilde{\rho}\right)}{\tilde{\rho}}\frac{nH_{n}^{\left(1\right)}\left(\tilde{\rho}'\right)}{\tilde{\rho}'}+r_{NN}\frac{\beta^{2}}{k_{2}^{2}}\partial H_{n}\left(\tilde{\rho}\right)\partial H_{n}\left(\tilde{\rho}'\right)\right.\\
 &  & \left.+r_{MN}\frac{\beta}{k_{2}}\left\{ \frac{nH_{n}\left(\tilde{\rho}\right)}{\tilde{\rho}}\partial H_{n}\left(\tilde{\rho}'\right)+\frac{nH_{n}^{\left(1\right)}\left(\tilde{\rho}'\right)}{\tilde{\rho}'}\partial H_{n}\left(\tilde{\rho}\right)\right\} \right]\\
\left[\overline{\overline{g}}_{n}\left(\tilde{\rho},\tilde{\rho}',\omega,\beta\right)\right]_{\rho\phi'} & = & r_{MM}\frac{nH_{n}^{\left(1\right)}\left(\tilde{\rho}'\right)}{\tilde{\rho}'}\partial H_{n}\left(\tilde{\rho}\right)+r_{NN}\frac{\beta^{2}}{k_{2}^{2}}\frac{nH_{n}^{\left(1\right)}\left(\tilde{\rho}'\right)}{\tilde{\rho}'}\partial H_{n}^{\left(1\right)}\left(\tilde{\rho}\right)\\
 &  & +r_{MN}\frac{\beta}{k_{2}}\left[\partial H_{n}^{\left(1\right)}\left(\tilde{\rho}\right)\partial H_{n}^{\left(1\right)}\left(\tilde{\rho}'\right)+\frac{nH_{n}^{\left(1\right)}\left(\tilde{\rho}\right)}{\tilde{\rho}}\frac{nH_{n}^{\left(1\right)}\left(\tilde{\rho}'\right)}{\tilde{\rho}'}\right]\\
\left[\overline{\overline{g}}_{n}\left(\tilde{\rho},\tilde{\rho}',\omega,\beta\right)\right]_{\rho z} & = & -r_{NM}\frac{nH_{n}^{\left(1\right)}\left(\tilde{\rho}\right)}{k_{2}\rho}H_{n}^{\left(1\right)}\left(\tilde{\rho}'\right)-r_{NN}\frac{\eta_{2}\beta}{k_{2}^{2}}H_{n}^{\left(1\right)}\left(\tilde{\rho}'\right)\partial H_{n}^{\left(1\right)}\left(\tilde{\rho}\right)\\
\left[\overline{\overline{g}}_{n}\left(\tilde{\rho},\tilde{\rho}',\omega,\beta\right)\right]_{\phi\phi'} & = & \mathrm{i}\left[r_{MM}\partial H_{n}^{\left(1\right)}\left(\tilde{\rho}\right)\partial H_{n}^{\left(1\right)}\left(\tilde{\rho}'\right)+r_{NN}\frac{\beta^{2}}{k_{2}^{2}}\frac{nH_{n}^{\left(1\right)}\left(\tilde{\rho}\right)}{\tilde{\rho}}\frac{nH_{n}^{\left(1\right)}\left(\tilde{\rho}'\right)}{\tilde{\rho}'}\right.\\
 &  & \left.+r_{MN}\frac{\beta}{k_{2}}\left\{ \frac{nH_{n}^{\left(1\right)}\left(\tilde{\rho}\right)}{\tilde{\rho}}\partial H_{n}^{\left(1\right)}\left(\tilde{\rho}'\right)+\frac{nH_{n}^{\left(1\right)}\left(\tilde{\rho}'\right)}{\tilde{\rho}'}\partial H_{n}^{\left(1\right)}\left(\tilde{\rho}\right)\right\} \right]\\
\left[\overline{\overline{g}}_{n}\left(\tilde{\rho},\tilde{\rho}',\omega,\beta\right)\right]_{\phi z} & = & -\mathrm{i}\left[r_{MN}\frac{\eta_{2}}{k_{2}}H_{n}^{\left(1\right)}\left(\tilde{\rho}'\right)\partial H_{n}^{\left(1\right)}\left(\tilde{\rho}\right)+r_{NN}\frac{\beta}{k_{2}}\frac{nH_{n}^{\left(1\right)}\left(\tilde{\rho}\right)}{k_{2}\rho}H_{n}^{\left(1\right)}\left(\tilde{\rho}'\right)\right]\\
\left[\overline{\overline{g}}_{n}\left(\tilde{\rho},\tilde{\rho}',\omega,\beta\right)\right]_{zz} & = & \mathrm{i}r_{NN}\frac{\eta_{2}^{2}}{k_{2}^{2}}H_{n}^{\left(1\right)}\left(\tilde{\rho}\right)H_{n}^{\left(1\right)}\left(\tilde{\rho}'\right)
\end{eqnarray*}
where we introduced $\partial H_{n}^{\left(1\right)}\left(x\right)\equiv\frac{dH_{n}^{\left(1\right)}\left(x\right)}{dx}$
and the reflection coefficients, $r_{MM}$, $r_{NN}$, and $r_{MN}=r_{NM}$,
defined by

\begin{eqnarray*}
r_{MM} & = & \frac{1}{D}\frac{J_{n}\left(\eta_{2}a\right)}{H_{n}^{\left(1\right)}\left(\eta_{2}a\right)}\left[\left(\frac{\beta n}{a}\right)^{2}\left(\frac{1}{\eta_{2}^{2}}-\frac{1}{\eta_{1}^{2}}\right)^{2}\right.\\
 &  & \left.-\left(\frac{\partial J_{n}\left(\eta_{1}a\right)}{\eta_{1}J_{n}\left(\eta_{1}a\right)}-\frac{\partial J_{n}\left(\eta_{2}a\right)}{\eta_{2}J_{n}\left(\eta_{2}a\right)}\right)\left(\frac{\partial J_{n}\left(\eta_{1}a\right)}{\eta_{1}J_{n}\left(\eta_{1}a\right)}k_{1}^{2}-\frac{\partial H_{n}^{\left(1\right)}\left(\eta_{2}a\right)}{\eta_{2}H_{n}^{\left(1\right)}\left(\eta_{2}a\right)}k_{2}^{2}\right)\right]\\
r_{NN} & = & \frac{1}{D}\frac{J_{n}\left(\eta_{2}a\right)}{H_{n}^{\left(1\right)}\left(\eta_{2}a\right)}\left[\left(\frac{\beta n}{a}\right)^{2}\left(\frac{1}{\eta_{2}^{2}}-\frac{1}{\eta_{1}^{2}}\right)^{2}\right.\\
 &  & \left.-\left(\frac{\partial J_{n}\left(\eta_{1}a\right)}{\eta_{1}J_{n}\left(\eta_{1}a\right)}k_{1}^{2}-\frac{\partial J_{n}\left(\eta_{2}a\right)}{\eta_{2}J_{n}\left(\eta_{2}a\right)}k_{2}^{2}\right)\left(\frac{\partial J_{n}\left(\eta_{1}a\right)}{\eta_{1}J_{n}\left(\eta_{1}a\right)}-\frac{\partial H_{n}^{\left(1\right)}\left(\eta_{2}a\right)}{\eta_{2}H_{n}^{\left(1\right)}\left(\eta_{2}a\right)}\right)\right]\\
r_{NM} & = & \frac{1}{D}\frac{k_{2}}{\eta_{2}}\left(\frac{\beta n}{a}\right)\frac{J_{n}\left(\eta_{2}a\right)}{H_{n}^{\left(1\right)}\left(\eta_{2}a\right)}\left(\frac{1}{\eta_{2}^{2}}-\frac{1}{\eta_{1}^{2}}\right)\left(\frac{\partial J_{n}\left(\eta_{2}a\right)}{J_{n}\left(\eta_{2}a\right)}-\frac{\partial H_{n}^{\left(1\right)}\left(\eta_{2}a\right)}{H_{n}^{\left(1\right)}\left(\eta_{2}a\right)}\right)
\end{eqnarray*}
with $D\equiv-\left(\frac{\beta n}{a}\right)^{2}\left(\frac{1}{\eta_{2}^{2}}-\frac{1}{\eta_{1}^{2}}\right)^{2}+\left(\frac{\partial J_{n}\left(\eta_{1}a\right)}{\eta_{1}J_{n}\left(\eta_{1}a\right)}-\frac{\partial H_{n}^{\left(1\right)}\left(\eta_{2}a\right)}{\eta_{2}H_{n}^{\left(1\right)}\left(\eta_{2}a\right)}\right)\left(\frac{\partial J_{n}\left(\eta_{1}a\right)}{\eta_{1}J_{n}\left(\eta_{1}a\right)}k_{1}^{2}-\frac{\partial H_{n}^{\left(1\right)}\left(\eta_{2}a\right)}{\eta_{2}H_{n}^{\left(1\right)}\left(\eta_{2}a\right)}k_{2}^{2}\right)$.
Note that $D$ and the reflection coefficients, $r_{AB}$, depend
on $n$, $\omega$, $a$, and $\beta$, i.e., $D=D_{n}\left(\omega,a,\beta\right)$
and $r_{AB}=r_{AB,n}\left(\omega,a,\beta\right)$. For the sake of
legibility, we omitted the index $n$ and arguments $\left(\omega,a,\beta\right)$
in the expressions above.

The contributions $\left[\overline{\overline{G}}_{\mathrm{sc}}\right]_{\phi\rho'}$,
$\left[\overline{\overline{G}}_{\mathrm{sc}}\right]_{z\rho'}$ and
$\left[\overline{\overline{G}}_{\mathrm{sc}}\right]_{z\phi'}$ can
be deduced from the previous expressions via the relation $\overline{\overline{G}}\left(\vec{r},\vec{r}',\omega\right)=\overline{\overline{G}}^{T}\left(\vec{r}',\vec{r},\omega\right)$.
We, moreover, note the following useful symmetry properties

\begin{eqnarray*}
\left[\overline{\overline{g}}_{n}\left(\tilde{\rho},\tilde{\rho}',\omega,-\beta\right)\right]_{ii'} & = & \left[\overline{\overline{g}}_{n}\left(\tilde{\rho},\tilde{\rho}',\omega,\beta\right)\right]_{ii'}\\
\left[\overline{\overline{g}}_{-n}\left(\tilde{\rho},\tilde{\rho}',\omega,\beta\right)\right]_{ii'} & = & \left[\overline{\overline{g}}_{n}\left(\tilde{\rho},\tilde{\rho}',\omega,\beta\right)\right]_{ii'}\\
\left[\overline{\overline{g}}_{-n}\left(\tilde{\rho},\tilde{\rho}',\omega,\beta\right)\right]_{\rho\phi'} & = & -\left[\overline{\overline{g}}_{n}\left(\tilde{\rho},\tilde{\rho}',\omega,\beta\right)\right]_{\rho\phi'}\\
\left[\overline{\overline{g}}_{n}\left(\tilde{\rho},\tilde{\rho}',\omega,-\beta\right)\right]_{\rho\phi'} & = & \left[\overline{\overline{g}}_{n}\left(\tilde{\rho},\tilde{\rho}',\omega,\beta\right)\right]_{\rho\phi'}\\
\left[\overline{\overline{g}}_{-n}\left(\tilde{\rho},\tilde{\rho}',\omega,\beta\right)\right]_{\rho z} & = & \left[\overline{\overline{g}}_{n}\left(\tilde{\rho},\tilde{\rho}',\omega,\beta\right)\right]_{\rho z}\\
\left[\overline{\overline{g}}_{n}\left(\tilde{\rho},\tilde{\rho}',\omega,-\beta\right)\right]_{\rho z} & = & -\left[\overline{\overline{g}}_{n}\left(\tilde{\rho},\tilde{\rho}',\omega,\beta\right)\right]_{\rho z}\\
\left[\overline{\overline{g}}_{-n}\left(\tilde{\rho},\tilde{\rho}',\omega,\beta\right)\right]_{\phi z} & = & -\left[\overline{\overline{g}}_{n}\left(\tilde{\rho},\tilde{\rho}',\omega,\beta\right)\right]_{\phi z}\\
\left[\overline{\overline{g}}_{n}\left(\tilde{\rho},\tilde{\rho}',\omega,-\beta\right)\right]_{\phi z} & = & -\left[\overline{\overline{g}}_{n}\left(\tilde{\rho},\tilde{\rho}',\omega,\beta\right)\right]_{\phi z}.
\end{eqnarray*}
In particular, these relations imply the scattering component, $\left.\overline{\overline{G}}_{\mathrm{sc}}\left(\vec{r},\vec{r}',\omega\right)\right|_{\vec{r}'=\vec{r}}$,
is diagonal in the $\left(\vec{e}_{\rho},\vec{e}_{\phi},\vec{e}_{z}\right)$
basis.

The poles of the integrand in Eq. (\ref{Gsc}) are found through solving
the equation $D_{n}\left[\omega,a,\beta\right]=0$ for $\beta$. The
pole equation coincides with the so-called characteristic equation
for the guided modes of a circular fiber. Such modes are fully determined
by a set $\mu\equiv\left(\text{K}_{lm},\omega,f,p\right)$ where $\text{K}=\text{TE}$,
TM (for $n=0$), HE, EH (for $n\neq0$) denotes the mode type, $p=\text{sign}\left(n\right)$,
$f=\pm1$, and the integers $l=\left|n\right|$ and $m$ are the azimuthal
and radial mode orders, respectively. The introduction of $f$ allows
one to consider only positive values for $\beta$. Indeed, by symmetry
of the characteristic equation, if $D_{n}\left[\omega,a,\beta\right]=0$,
then $D_{n}\left[\omega,a,-\beta\right]=0$. By convention, the value
of $\beta$ for the mode $\mu=\left(\text{K}_{lm},\omega,f=+1,p\right)$,
denoted by $\beta_{\mu}\left(a\right)$, is chosen positive, while
the value of $\beta$ for the mode $\mu=\left(\text{K}_{lm},\omega,f=-1,p\right)$
is $-\beta_{\mu}\left(a\right)<0$. With these definitions, we apply
the residue theorem to Eq. (\ref{Gsc}) and get the following decomposition
\citep{KD04,AMA17}

\begin{eqnarray*}
\overline{\overline{G}}_{\mathrm{sc}}\left(\vec{r},\vec{r},\omega\right) & = & \overline{\overline{G}}_{\mathrm{r}}\left(\vec{r},\vec{r},\omega\right)+\overline{\overline{G}}_{\mathrm{g}}\left(\vec{r},\vec{r},\omega\right)\\
\overline{\overline{G}}_{\mathrm{r}}\left(\vec{r},\vec{r},\omega\right) & = & \frac{1}{8\pi}\sum_{n=-\infty}^{+\infty}\intop_{-\omega/c}^{\omega/c}\mathrm{d}\beta\;\overline{\overline{g}}_{n}\left(\tilde{\rho},\tilde{\rho},\omega,\beta\right)\\
\overline{\overline{G}}_{\mathrm{g}}\left(\vec{r},\vec{r},\omega\right) & = & \frac{\mathrm{i}}{4\pi}\sum_{\text{K}=\text{TE},\text{TM}}\sum_{f=\pm1}\sum_{m}\mathrm{Res}\left[\overline{\overline{g}}_{0}\left(\tilde{\rho},\tilde{\rho},\omega,f\beta_{\text{K}_{0m}}\right)\right]\\
 &  & +\frac{\mathrm{i}}{4\pi}\sum_{l=1}^{+\infty}\sum_{\text{K}=\text{HE},\text{EH}}\sum_{f,p=\pm1}\sum_{m}\mathrm{Res}\left[\overline{\overline{g}}_{pl}\left(\tilde{\rho},\tilde{\rho},\omega,f\beta_{\text{K}_{lm}}\right)\right]
\end{eqnarray*}
where $\overline{\overline{G}}_{\mathrm{r}}$ and $\overline{\overline{G}}_{\mathrm{g}}$
are interpreted as the contributions of radiative modes 
\[
\sigma=\left(\omega,\beta\in\left[-\nicefrac{\omega}{c},\nicefrac{\omega}{c}\right],n=\cdots-1,0,1\cdots,p=\pm1\right)
\]
and guided modes $\mu=\left(\text{K}_{lm},\omega,f,p\right)$, respectively.
Following the analogy with the electromagnetic wave theory of fiber
modes, we identify $\beta$ with the propagation constant, i.e., the
projection $k_{z}$ of the mode wavevector onto the fiber axis, $\left(Oz\right)$.
To be more explicit, for radiative modes $\left(\sigma\right)$ $k_{\sigma,z}=\beta$,
while for guided modes $\left(\mu\right)$ $k_{\mu,z}=f\beta_{\mu}$.

\section{Force and anisotropy\label{AppF}}

The Lorentz force on an atom located at a position, $\vec{R}$, in
an electromagnetic field $\left(\vec{E},\vec{B}\right)$ takes the
form

\begin{eqnarray*}
\vec{F}\left(t\right) & = & \vec{\nabla}\langle\hat{\vec{d}}\cdot\hat{\vec{E}}\left(\vec{r},t\right)\rangle|_{\vec{r}=\vec{R}}+\frac{\mathrm{d}}{\mathrm{d}t}\langle\hat{\vec{d}}\times\hat{\vec{B}}\left(\vec{r},t\right)\rangle|_{\vec{r}=\vec{R}}
\end{eqnarray*}
Assuming the atom is initially in a statistical mixture of states
$\left\{ \left|n\right\rangle \right\} $, the general expression
of this force is \citep{Buh12}

\begin{eqnarray}
\vec{F}\left(t\right) & = & \sum_{n}p_{n}\left(t\right)\vec{F}_{n}\nonumber \\
\vec{F}_{n} & = & \sum_{k}\frac{\mu_{0}}{\pi}\intop_{0}^{+\infty}\mathrm{d}\omega~\omega^{2}\frac{\vec{\nabla}_{\vec{r}}\left[\vec{d}_{nk}\cdot\mathrm{Im}\left[\overline{\overline{G}}_{\text{sc}}\left(\vec{r},\vec{R},\omega\right)\right]\cdot\vec{d}_{kn}\right]|_{\vec{r}=\vec{R}}}{\omega-\omega_{nk}-\frac{\mathrm{i}}{2}\left(\Gamma_{n}+\Gamma_{k}\right)}+\text{h.c.}\label{EQFn}
\end{eqnarray}
where $\Gamma_{n}$ is the spontaneous emission from the excited state
$\left|n\right\rangle $, $p_{n}\left(t\right)$ is the population
of state $\left|n\right\rangle $ at time $t$, $\vec{d}_{nk}\equiv\langle n|\hat{\vec{d}}|k\rangle$.
We neglect broadening in the denominator of the integrand in Eq. (\ref{EQFn}),
i.e., $\omega_{kn}+\frac{\mathrm{i}}{2}\left(\Gamma_{n}+\Gamma_{k}\right)\approx\omega_{kn}$.
Then, by application of the residue theorem, we split this force into
a resonant and a nonresonant part, i.e., $\vec{F}_{n}=\vec{F}_{n}^{\text{res}}+\vec{F}_{n}^{\text{nres}}$,
with

\begin{eqnarray*}
\vec{F}_{n}^{\mathrm{res}} & = & \sum_{k<n}2\mu_{0}\omega_{nk}^{2}\mathrm{Re}\left(\vec{\nabla}_{\vec{r}}\left.\left[\vec{d}_{nk}\cdot\overline{\overline{G}}_{\text{sc}}\left(\vec{r},\vec{R},\omega_{nk}\right)\cdot\vec{d}_{kn}\right]\right|_{\vec{r}=\vec{R}}\right)\\
\vec{F}_{n}^{\text{nres}} & = & -\frac{\mu_{0}}{\pi}\intop_{0}^{+\infty}\mathrm{d}\xi~\xi^{2}\frac{\omega_{kn}}{\omega_{kn}^{2}+\xi^{2}}\nabla_{\vec{r}}\left[\vec{d}_{nk}\cdot\overline{\overline{G}}_{\text{sc}}\left(\vec{r},\vec{R},\mathrm{i}\xi\right)|_{\vec{r}=\vec{R}}\cdot\vec{d}_{kn}\right].
\end{eqnarray*}

\noindent We emphasize that the nonresonant part is summed over all
transitions, while the resonant part takes into account only radiative
transitions towards states $\left|k\right\rangle $ of lower energy
than $\left|n\right\rangle $. From the symmetry properties of $\overline{\overline{g}}_{n}$,
one deduces

\begin{eqnarray*}
\left[\frac{\partial}{\partial z}\overline{\overline{G}}_{\text{sc}}\left(\vec{r},\vec{R}\right)|_{\vec{r}=\vec{R}}\right]_{ii} & = & \left[\frac{1}{R}\frac{\partial}{\partial\phi}\overline{\overline{G}}_{\text{sc}}\left(\vec{r},\vec{R}\right)|_{\vec{r}=\vec{R}}\right]_{ii}=0\\
\left[\frac{\partial}{\partial z}\overline{\overline{G}}_{\text{sc}}\left(\vec{r},\vec{R}\right)|_{\vec{r}=\vec{R}}\right]_{\rho\phi} & = & \left[\frac{\partial}{\partial z}\overline{\overline{G}}_{\text{sc}}\left(\vec{r},\vec{R}\right)|_{\vec{r}=\vec{R}}\right]_{z\phi}=0\\
\left[\frac{\partial}{\partial\phi}\overline{\overline{G}}_{\text{sc}}\left(\vec{r},\vec{R}\right)|_{\vec{r}=\vec{R}}\right]_{\rho z} & = & \left[\frac{\partial}{\partial\phi}\overline{\overline{G}}_{\text{sc}}\left(\vec{r},\vec{R}\right)|_{\vec{r}=\vec{R}}\right]_{z\phi}=0
\end{eqnarray*}
Setting $\left[\overline{\overline{G}}_{\text{sc}}\left(\vec{r},\vec{R},\omega\right)\right]_{ii}\equiv G_{ii}$
and $\left[d_{nk}\right]_{i}\equiv d_{i}$ for shortness, one gets

\begin{eqnarray*}
\nabla_{\vec{r}}\left[\vec{d}_{nk}\cdot\overline{\overline{G}}_{\text{sc}}\left(\vec{r},\vec{R},\omega_{nk}\right)\cdot\vec{d}_{kn}\right]|_{\vec{r}=\vec{R}} & = & \frac{\partial}{\partial\rho}\left[\left|d_{\rho}\right|^{2}G_{\rho\rho}\left(\vec{r},\vec{R}\right)+\left|d_{\phi}\right|^{2}G_{\phi\phi}\left(\vec{r},\vec{R}\right)\right.\\
 &  & \left.+\left|d_{z}\right|^{2}G_{zz}\left(\vec{r},\vec{R}\right)\right]|_{\vec{r}=\vec{R}}\vec{e}_{x}\\
 & + & 2\mathrm{i}\mathrm{Im}\left(d_{\rho}d_{\phi}^{*}\right)\frac{1}{R}\frac{\partial}{\partial\phi}G_{\rho\phi}\left(\vec{r},\vec{R}\right)|_{\vec{r}=\vec{R}}\vec{e}_{y}\\
 & + & 2\mathrm{i}\mathrm{Im}\left(d_{\rho}d_{z}^{*}\right)\frac{\partial}{\partial z}G_{\rho z}\left(\vec{r},\vec{R}\right)|_{\vec{r}=\vec{R}}\vec{e}_{z}
\end{eqnarray*}
Finally, using $\vec{\nabla}_{\vec{r}}G_{ij}\left(\vec{r},\vec{R}\right)|_{\vec{r}=\vec{R}}=\frac{1}{2}\vec{\nabla}_{\vec{r}}G_{ij}\left(\vec{r},\vec{r}\right)|_{\vec{r}=\vec{R}}$
and noticing that $2\mathrm{Re}\left[\mathrm{i}\partial_{k}G_{ij}\right]=-2\mathrm{Im}\left[\partial_{k}G_{ij}\right]$,
we can get the resonant force projection in the $\left(\vec{e}_{x},\vec{e}_{y},\vec{e}_{z}\right)$
basis (which corresponds to the cylindrical basis $\left(\vec{e}_{\rho},\vec{e}_{\phi},\vec{e}_{z}\right)$
at the location of the atom, see Fig. \ref{System})

\begin{eqnarray*}
\left[F_{n}^{\textrm{res}}\right]_{x} & = & \sum_{k<n}\frac{\partial}{\partial\rho}\mathrm{Re}\left[\mu_{0}\omega_{nk}^{2}\vec{d}_{nk}\cdot\overline{\overline{G}}_{\text{sc}}\left(\vec{r},\vec{r},\omega_{nk}\right)\cdot\vec{d}_{kn}\right]|_{\vec{r}=\vec{R}}\\
\left[F_{n}^{\textrm{res}}\right]_{y} & = & -\sum_{k<n}4\mu_{0}\omega_{kn}^{2}\mathrm{Im}\left(d_{x}d_{y}^{*}\right)\mathrm{Im}\left[\frac{1}{R}\frac{\partial}{\partial\phi}G_{xy}\left(\vec{r},\vec{r}\right)|_{\vec{r}=\vec{R}}\right]\\
\left[F_{n}^{\textrm{res}}\right]_{z} & = & -\sum_{k<n}4\mu_{0}\omega_{kn}^{2}\mathrm{Im}\left(d_{x}d_{z}^{*}\right)\mathrm{Im}\left[\frac{\partial}{\partial z}G_{xz}\left(\vec{r},\vec{r}\right)|_{\vec{r}=\vec{R}}\right]
\end{eqnarray*}
and the nonresonant projection~:

\begin{eqnarray*}
\left[F_{n}^{\mathrm{nres}}\right]_{x} & = & -\sum_{k}\frac{\partial}{\partial\rho}\left(\frac{\mu_{0}}{\pi}\intop_{0}^{+\infty}\mathrm{d}\xi~\frac{\xi^{2}\tilde{\omega}_{kn}}{\tilde{\omega}_{kn}^{2}+\xi^{2}}\vec{d}_{nk}\cdot\overline{\overline{G}}_{\text{sc}}\left(\vec{r},\vec{r},\mathrm{i}\xi\right)\cdot\vec{d}_{kn}\right)|_{\vec{r}=\vec{R}}.\\
\left[F_{n}^{\mathrm{nres}}\right]_{y} & = & \frac{4\mu_{0}}{\pi}\sum_{k}\mathrm{Im}\left[d_{x}d_{y}^{*}\right]\intop_{0}^{+\infty}\mathrm{d}\xi~\frac{\xi^{2}\tilde{\omega}_{kn}}{\tilde{\omega}_{kn}^{2}+\xi^{2}}\frac{1}{R}\frac{\partial}{\partial\left(\mathrm{i}\phi\right)}G_{xy}\left(\vec{r},\vec{r},\mathrm{i}\xi\right)|_{\vec{r}=\vec{R}}\\
\left[F_{n}^{\mathrm{nres}}\right]_{z} & = & \frac{4\mu_{0}}{\pi}\sum_{k}\mathrm{Im}\left[d_{x}d_{z}^{*}\right]\intop_{0}^{+\infty}\mathrm{d}\xi~\frac{\xi^{2}\tilde{\omega}_{kn}}{\tilde{\omega}_{kn}^{2}+\xi^{2}}\frac{\partial}{\partial\left(\mathrm{i}z\right)}G_{xz}\left(\vec{r},\vec{r},\mathrm{i}\xi\right)|_{\vec{r}=\vec{R}}
\end{eqnarray*}

The radial component (i.e., along $x$) can be expressed as the derivative
of the energy displacement, i.e., $F_{x}=F_{x}^{\mathrm{nres}}+F_{x}^{\mathrm{res}}=-\frac{\partial}{\partial\rho}\left[\hbar\delta\omega^{\mathrm{nres}}\left(\rho\right)+\hbar\delta\omega^{\mathrm{res}}\left(\rho\right)\right]|_{\vec{r}=\vec{R}}$.
This result justifies the Casimir-Polder approach in which the radial
force derives from the potentiel $U\left(\rho\right)=\hbar\delta\omega\left(\rho\right)$
related to the energy displacement.

The resonant forces along $y$ and $z$ can be interpreted as resulting
from average recoil forces due to the preferential emission of photons
of a given polarization $\left(\left[F_{n}^{\text{res}}\right]_{y}\right)$
or towards a given direction $\left(\left[F_{n}^{\text{res}}\right]_{z}\right)$.
Using the results of the previous Appendix, one can moreover decompose
these forces as sums of the contributions of the different modes and
atomic transitions : e.g. $\left[F_{n}^{\text{res}}\right]_{z}=\sum_{\nu,k<n}\left[F_{nk,\nu}^{\text{res}}\right]_{z}$
where $\left[F_{nk,\nu}^{\text{res}}\right]_{z}$ is the force relative
to the transition $\left|n\right\rangle \rightarrow\left|k\right\rangle $
coupled to the (guided or radiative) mode $\nu$.

\section{Electric dipole and quadrupole transitions\label{AppEDT}}

The electric dipole and quadrupole contributions to the interaction
Hamiltonian of an atom located at position $\vec{R}$ with the electromagnetic
field can be written as

\begin{eqnarray*}
\hat{H}_{\mathrm{dip}} & = & -\vec{d}\cdot\vec{E}\left(\vec{R}\right)\\
\hat{H}_{\mathrm{quad}} & = & -\overline{\overline{Q}}\bullet\left[\vec{\nabla}\otimes\hat{\vec{E}}\left(\vec{R}\right)\right]
\end{eqnarray*}
where $\bullet$ denotes the Frobenius inner product explicitly defined
by $\overline{\overline{A}}\bullet\overline{\overline{B}}=\sum_{i,j}A_{ij}B_{ji}$,
$\left\{ A_{ij}\right\} $ being the components of the tensor $\overline{\overline{A}}$
in an orthonormal basis \citep{Buh12}, and
\begin{eqnarray*}
\vec{d} & = & er\vec{e}_{r}\\
\overline{\overline{Q}} & = & \frac{e}{2}r^{2}\vec{e}_{r}\otimes\vec{e}_{r}
\end{eqnarray*}
In the dipole and quadrupole operators above, $r\vec{e}_{r}$ (approximately)
corresponds to the position of the active valence electron with respect
to the nucleus of the atom. The matrix elements $\vec{d}_{nk}\equiv\left\langle n\left|\vec{d}\right|k\right\rangle $
and $\overline{\overline{Q}}_{nk}=\langle n|\overline{\overline{Q}}|k\rangle$
comprise radial and angular parts. The radial parts $\langle n'l'j'|\hat{r}|nlj\rangle$
and $\langle n'l'j'|\hat{r}^{2}|nlj\rangle$ can be computed thanks
to the Alkali Rydberg Calculator \citep{ARC}. To get the angular
parts, we express $\vec{e}_{r}$ and $\vec{e}_{r}\otimes\vec{e}_{r}$
in the basis $\left(\vec{e}_{x},\vec{e}_{y},\vec{e}_{z}\right)$ in
terms of spherical harmonics $Y_{l,q}$

\begin{eqnarray*}
\vec{e}_{r} & = & \sqrt{\frac{2\pi}{3}}\left(\begin{array}{c}
Y_{1,-1}-Y_{1,1}\\
\mathrm{i}\left(Y_{1,-1}+Y_{1,1}\right)\\
\sqrt{2}Y_{1,0}
\end{array}\right)\\
\vec{e}_{r}\otimes\vec{e}_{r} & = & \sqrt{\frac{\pi}{30}}\left(\begin{array}{c}
\left(Y_{2,-2}+Y_{2,2}\right)-\sqrt{\frac{2}{3}}Y_{2,0}+\sqrt{\frac{10}{3}}Y_{0,0}\\
\mathrm{i}\left(Y_{2,-2}-Y_{2,2}\right)\\
Y_{2,-1}-Y_{2,1}
\end{array}\right.\\
 &  & \left.\begin{array}{cc}
\mathrm{i}\left(Y_{2,-2}-Y_{2,2}\right) & Y_{2,-1}-Y_{2,1}\\
-\left(Y_{2,-2}+Y_{2,2}\right)-\sqrt{\frac{2}{3}}Y_{2,0}+\sqrt{\frac{10}{3}}Y_{0,0} & \mathrm{i}\left(Y_{2,-1}+Y_{2,1}\right)\\
\mathrm{i}\left(Y_{2,-1}+Y_{2,1}\right) & \sqrt{\frac{8}{3}}Y_{2,0}+\sqrt{\frac{10}{3}}Y_{0,0}
\end{array}\right)
\end{eqnarray*}
and use the following formula \citep{Sob12}

\begin{eqnarray*}
\left\langle l,j,F,M\left|Y_{k,q}\right|l',j',F',M\right\rangle \\
=\left(-1\right)^{j+j'+I+s+k-M}\sqrt{\frac{1}{4\pi}\left(2k+1\right)\left(2l+1\right)\left(2l'+1\right)\left(2J+1\right)\left(2J'+1\right)\left(2F+1\right)\left(2F'+1\right)}\\
\times\left(\begin{array}{ccc}
l & k & l'\\
0 & 0 & 0
\end{array}\right)\left\{ \begin{array}{ccc}
l & j & s\\
j' & l' & k
\end{array}\right\} \left\{ \begin{array}{ccc}
j & F & I\\
F' & j' & k
\end{array}\right\} \left(\begin{array}{ccc}
F' & k & F\\
M' & q & -M
\end{array}\right)
\end{eqnarray*}
Finally, we can compute the spontaneous emission rates along the transition
$|n\rangle\rightarrow|k\rangle$ due to dipole and quadrupole terms,
respectively, to be given by

\begin{eqnarray*}
\Gamma_{nk}^{\left(\mathrm{dip}\right)} & = & \frac{2\mu_{0}}{\hbar}\omega_{nk}^{2}\sum_{\alpha,\beta=x,y,z}\left[d_{nk}\right]_{\alpha}\left[d_{kn}\right]_{\beta}\mathrm{Im}\left[\overline{\overline{G}}_{\alpha\beta}\left(\vec{R},\vec{R}',\omega_{nk}\right)\right]\\
\Gamma_{nk}^{\left(\mathrm{quad}\right)} & = & \lim_{\left|\vec{R}-\vec{R}'\right|\rightarrow0}\frac{2\mu_{0}}{\hbar}\omega_{nk}^{2}\sum_{\alpha,\beta=x,y,z}\left[\overline{\overline{Q}}_{nk}\right]_{\alpha\beta}\left[\overline{\overline{Q}}_{kn}\right]_{\gamma\delta}\partial_{\alpha}\partial'_{\gamma}\mathrm{Im}\left[\overline{\overline{G}}_{\beta\delta}\left(\vec{R},\vec{R}',\omega_{nk}\right)\right]
\end{eqnarray*}
and the van der Waals potential in the non-retarded approximation
is given by

\begin{eqnarray*}
U_{n}^{\left(\mathrm{dip}\right)}\left(\vec{R}\right) & = & -\frac{1}{2\epsilon_{0}}\sum_{\alpha,\beta=x,y,z}\sum_{k}\left[d_{nk}\right]_{\alpha}\left[d_{kn}\right]_{\beta}\left[\overline{\overline{\Gamma}}_{0}\left(\vec{R}\right)\right]_{\alpha\beta}\\
U_{n}^{\left(\mathrm{quad}\right)}\left(\vec{R}\right) & = & -\frac{1}{2\epsilon_{0}}\sum_{\alpha,\beta=x,y,z}\sum_{k}\left[Q_{nk}\right]_{\alpha\beta}\left[Q_{kn}\right]_{\gamma\delta}\partial_{\alpha}\partial'_{\gamma}\left[\overline{\overline{\Gamma}}_{0}\left(\vec{R}\right)\right]_{\beta\delta}
\end{eqnarray*}
where we introduced $\overline{\overline{\Gamma}}_{0}\left(\vec{R}\right)\equiv\lim_{\omega\rightarrow0}\;\frac{\omega^{2}}{c^{2}}\overline{\overline{G}}\left(\vec{R},\vec{R},\omega\right)$.

\end{document}